\documentclass{sig-alternate-05-2015}
\usepackage{subfigure}
\usepackage{epstopdf}
\usepackage{color}
\usepackage{graphicx}
\usepackage{balance}  
\usepackage{times,amsmath,epsfig,cite,amssymb,amsfonts,url,dsfont,multirow,hyperref}
\usepackage{algorithm}
\usepackage{algorithmic}
\usepackage{url}
\usepackage{bbm}
\usepackage{multirow}
\usepackage{ntheorem}   
\usepackage{mdframed}   
\usepackage{blkarray}
\DeclareMathOperator*{\argmin}{arg\,min}

\newcommand{\eat}[1]{}

\newdef{definition}{Definition}
\newtheorem{problem} {Problem}

\definecolor{light-gray}{gray}{0.95}

\newmdtheoremenv[%
leftmargin=0pt,%
rightmargin=0pt,backgroundcolor=light-gray,
innertopmargin=0.5\topskip,%
splittopskip=\topskip,skipbelow=\baselineskip,%
skipabove=\baselineskip,ntheorem]{myans}{Result}

\newfont{\mycrnotice}{ptmr8t at 7pt}
\newfont{\myconfname}{ptmri8t at 7pt}
\let\crnotice\mycrnotice%
\let\confname\myconfname%

\begin{document}
%
\title{Latent Space Model for Road Networks to Predict Time-Varying Traffic}
\author{
	%
	%
	Dingxiong Deng$^1$, Cyrus Shahabi$^1$, Ugur Demiryurek$^1$, Linhong Zhu$^2$,  Rose Yu$^1$, Yan Liu$^1$\\
	\affaddr{Department of Computer Science, University of Southern California$^1$}\\
	\affaddr{Information Sciences Institute, University of Southern California$^2$}\\
	\email{\{dingxiod, shahabi, demiryur, qiyu, Yanliu.cs\}@usc.edu, linhong@isi.edu}
}

\maketitle
\begin{abstract}
Real-time traffic prediction from high-fidelity spatiotemporal traffic sensor datasets is an important problem for intelligent transportation systems and sustainability. However, it is challenging due to the complex topological dependencies and high dynamism associated with changing road conditions. In this paper, we propose a Latent Space Model for Road Networks (LSM-RN) to address these challenges holistically. In particular, given a series of road network snapshots, we learn the attributes of vertices in latent spaces which capture both topological and temporal properties. As these latent attributes are time-dependent, they can estimate how traffic patterns form and evolve. In addition, we present an incremental online algorithm which sequentially and adaptively learns the latent attributes from the temporal graph changes. Our framework enables real-time traffic prediction by 1) exploiting real-time sensor readings to adjust/update the existing latent spaces, and 2) training as data arrives and making predictions on-the-fly. By conducting extensive experiments with a large volume of real-world traffic sensor data, we demonstrate the superiority of our framework for real-time traffic prediction on large road networks over competitors as well as baseline graph-based LSM's.

\end{abstract}

%
%

\section{Introduction}\label{sec-intro}
Recent advances in traffic sensing technology have enabled the acquisition  of high-fidelity spatiotemporal traffic datasets. For example, at our research center, for the past five years, we have been collecting data from 15000 loop detectors installed on the highways and arterial streets of Los Angeles County, covering 3420 miles cumulatively (see the case study in~\cite{jagadish2014big}). The collected data include several main traffic parameters such as occupancy, volume, and speed at the rate of 1 reading/sensor/min. These data streams enable traffic prediction, which in turn improves route navigation, traffic regulation, urban planning, etc.

The \textit{traffic prediction} problem is to predict the future travel speed of each and every edge of a road network, given the historical speed readings sensed from the sensors on these edges. To solve the traffic prediction problem, the majority of existing techniques utilize the historical information of an edge to predict its future travel-speed using regression techniques such as Auto-regressive Integrated Moving Average (ARIMA)~\cite{PanICDM12}, Support Vector Regression (SVR)~\cite{RistPAKDD13} and Gaussian Process (GP)~\cite{ZhouSigmod15}. There are also other studies that leverage spatial/topological similarities to predict the readings of an edge based on its neighbors in either Euclidean space~\cite{haworth2012non} or network~\cite{kwon2000modeling}. Even though there are few notable exceptions such as Hidden Markov Model (HMM)~\cite{kwon2000modeling,yang2013travel} that predict traffic of edges by collectively inferring temporal information, these approaches simply combine the local information of neighbors with temporal information. Furthermore, existing approaches such as GP and HMM are computationally expensive and require repeated offline trainings. Therefore, it is very difficult to adapt these to real-time traffic forecasting.

Motivated by these challenges, we propose Latent Space Modeling for Road Networks (\textbf{LSM-RN}), which enables more accurate and scalable traffic prediction by utilizing both topology similarity and temporal correlations. Specifically, with LSM-RN, vertices of dynamic road network are embedded into a latent space, where two vertices that are similar in terms of both time-series traffic behavior and the road network topology are close to each other in the latent space. Recently, several machine learning problems such as community detection~\cite{zhang2012overlapping,wang2011community}, link prediction~\cite{menon2011link,corrZhuSG14} and sentimental analysis~\cite{zhu2014tripartite} have been formulated using Latent Space Modeling. Most relevant to our work is the latent space modeling for social networks (hereafter called \textbf{LSM-SN}) because RN and SN are both graphs and each vertex has different attributes. However, existing approaches for LSM-SN are not suitable for both identifying the edge and/or sensor latent attributes in road networks and exploiting them for real-time traffic prediction due to the following reasons.

First, road networks show significant topological(e.g., travel-speeds - weights - between two sensors on the same road segment are similar), and temporal (e.g., travel-speeds measured every 1 minute on a particular sensor are similar) correlations. These correlations can be exploited to alleviate the missing data problem, which is unique to road networks, due to the fact that some road segments may contain no sensors and any sensor may occasionally fail to report data. Second, unlike social networks, LSM-RN is fast evolving due to the time-varying traffic conditions. On the contrary, social networks evolve smoothly and frequent changes are very unlikely (e.g., one user changes its political preferences twice a day). Instead, in road networks, traffic conditions on a particular road segment can change rapidly in a short time (i.e., time-dependent) because of rush/non-rush hours and traffic incidents. Third, LSM-RN is highly dynamic where fresh data come in a streaming fashion, whereas the connection (weights) between nodes in social networks is almost static. The dynamism requires partial updates of the model as opposed to the time-consuming full updates in LSM-SN. Finally, with LSM-RN, the ground truth can be observed shortly after making the prediction (by measuring the actual speed later in future), which also provides an opportunity to improve/adjust the model incrementally (i.e., online learning).

With our proposed LSM-RN, each dimension of the embedded latent space represents a latent attribute, thus the attribute distribution of vertices and how the attributes interact with each other jointly determine the underlying traffic pattern. To enforce the topology of road network, LSM-RN adds a graph Laplacian constraint which not only enables global graph similarity, but also completes the missing data by a set of similar edges with non-zero readings. Subsequently, we incorporate the temporal properties into our LSM-RN model by considering time-dependent latent attributes and a global transition process. With these time-dependent latent attributes and the transition matrix, we are able to understand how traffic patterns form and evolve.

To infer the time-dependent latent attributes of our LSM-RN model, a typical method is to utilize multiplicative update algorithms~\cite{lee2001algorithms}, where we jointly infer the whole latent attributes via iterative updates until they become stable, termed as \emph{global learning}. However, global learning is not only slow but also not practical for real-time traffic prediction. This is because, the spatiotemporal traffic data are of high-fidelity (i.e., updates are frequent in every one minute) and  the actual ground-truth of traffic speed becomes available shortly afterwards ( e.g., after making a prediction for the next five minutes, the ground truth data will be available instantly after five minutes). We thus propose an incremental online learning with which we sequentially and adaptively learn the latent attribute from the temporal traffic changes. In particular, each time when our algorithm makes a prediction with the latent attributes learned from the previous snapshot, it receives feedback from the next snapshot (i.e., the ground truth speed reading we already obtained) and subsequently modifies the latent attributes for more accurate predictions. Unlike traditional online learning which only performs one single update (e.g., update one vertex per prediction) per round, our goal is to make predictions for the entire road network, and thus we update the latent attributes for many correlated vertices.

Leveraging global and incremental learning algorithms our LSM-RN model can strike a balance between accuracy and efficiency for real-time forecasting. Specifically, we consider a setting with a predefined time window where at each time window (e.g., 5 minutes), we learn our traffic model with the proposed incremental inference approach on-the-fly, and make predictions for the next time span. Meanwhile, we batch the re-computation of our traffic model at the end of one large time window (e.g., one hour). Under this setting, our LSM-RN model enables the following two properties: (1) real-time feedback information can be seamlessly incorporated into our framework to adjust for the existing latent spaces, thus allowing for more accurate predictions, and (2) our algorithms train and make predictions on-the-fly with small amount of data rather than requiring large training datasets.

We conducted extensive experiments using a large volume of real-world traffic sensor dataset. We demonstrated that the LSM-RN framework achieves better accuracy than that of both existing time series methods (e.g. ARIMA and SVR) and the LSM-SN approaches. Moreover, we show that our algorithm scales to large road networks. For example, it only takes 4 seconds to make a prediction for a network with 19,986 edges. Finally, we show that our batch window setting works perfectly for streaming data,
alternating the executions of our global and incremental algorithms, which strikes a compromise between prediction accuracy and efficiency. For instance, incremental learning is one order of magnitude faster than global learning, and it requires less than 1 seconds to incorporate real-time feedback information. 

The remainder of this paper is organized as follows. In Section~\ref{sec-related}, we discuss the related work. We define our problem in Section~\ref{sec-problem}, and explain LSM-RN in Section~\ref{sec-model}.  We present the global learning and increment learning algorithms, and  discuss how to adapt our algorithms for real-time traffic forecasting in Section~\ref{sec-global}.
In Section~\ref{sec-exp}, we report the experiment results and Section~\ref{sec-con} concludes the paper.

\vspace{-3mm}
\section{Background and Related Works}\label{sec-related}
\vspace{-3mm}
\subsection{Traffic analysis}
Many studies have focused on the traffic prediction problem,  but no single study so far tackled all the challenges in a holistic manner. Some focused on missing values~\cite{qu2008bpca} or missing sensors~\cite{ide2011trajectory,zheng2013time}, but not both. Some studies~\cite{ZhouSigmod15, PanICDM12} focus on utilizing temporal data which models each sensor (or edge) independently and makes predictions using time series approaches (e.g., Auto-regressive Integrated Moving Average (ARIMA)~\cite{PanICDM12}, Support Vector Regression (SVR)~\cite{RistPAKDD13} and Gaussian Process (GP)~\cite{ZhouSigmod15}). For instance, Pan et. al.~\cite{PanICDM12} learned an enhanced ARIMA model for each edge in advance, and then perform traffic prediction on top of these models. Very few studies~\cite{yang2013travel,kwon2000modeling} utilize spatiotemporal model with correlated time series based on Hidden Markov Model, but only for small number of time series and not always using the network space as the spatial dimension (e.g., using Euclidean space~\cite{haworth2012non}).
In~\cite{xu2015mining}, Xu et. al. consider using the newly arrived data as feedback to reward one classifier vs. the other but not for dynamically updating the model. Note that many existing studies~\cite{zheng2013time,ide2011trajectory,wang2014travel, yang2013travel,cheng2008semi} on traffic prediction are based on GPS dataset, which is different with the sensor dataset, where we have fine-grained and steady readings from road-equipped sensors. 
We are not aware of any study that uses latent space modeling (considering both time and network topology) for real-time traffic prediction from incomplete (i.e., missing sensors and values) sensor datasets.

\vspace{-2mm}
\subsection{Latent space model and NMF}
Recently, many real data analytic problems such as community detection~\cite{zhang2012overlapping,wang2011community}, recommendation system~\cite{chua2013modeling}, topic modeling~\cite{saha2012learning}, image clustering~\cite{CaiPAMI2011}, and sentiment analysis~\cite{zhu2014tripartite},  have been formulated as the problem of latent space learning. These studies assume that, given a graph, each vertex resides in a latent space with attributes, and vertices which are close to each other are more likely to be in the same cluster (e.g., community or topic) and form a link. In particular, the focus is to infer the latent matrix by minimizing the loss (e.g., squared loss~\cite{zhang2012overlapping, zhu2014tripartite} or KL-divergence~\cite{CaiPAMI2011}) between observed and estimated links. However, existing methods are not designed for the highly correlated (topologically and temporally) and very dynamic road networks. Few studies~\cite{rossi2013modeling} have considered the temporal relationships in SN with the assumption that networks are evolving smoothly. In addition, the temporal graph snapshots in~\cite{rossi2013modeling} are treated separately and thus newly observed data will not be incorporated to improve the model. Compared with existing works, we explore the feasibility of modeling road networks with time-varying latent space. The traffic speed of a road segment is determined by their latent attributes and the interaction between corresponding attributes. To tackle the sparsity of road network, we utilize the graph topology by adding a graph Laplacian constraint to impute the missing values. In addition, the latent position of each vertex, is varying with time and allows for sudden movement from one timestamp to the next timestamp via a transition matrix.

Different techniques have been proposed to learn the latent properties, where Non-negative Matrix Factorization (NMF) is one of the most popular method because of its ease of interpreting and flexibility. In this work, we explore the feasibility of applying dynamic NMF to traffic prediction domain. We design a global algorithm to infer the latent space based on the traditional multiplicative algorithm~\cite{lee2001algorithms, ding2006orthogonal}. We further propose a topology-aware incremental algorithm, which adaptively updates the latent space representation for each node in the road network with topology constraints. The proposed algorithms differ from traditional online NMF algorithms such as~\cite{blondel2014online}, which independently perform each online update.

\vspace{-4mm}
\begin{table}[!ht]
	\centering
	\caption{{\small Notations and explanations}}\label{tbl:notation}
	\begin{small}
		\scalebox{0.75}{
			\begin{tabular}{|l|l|}
				\hline
				Notations&Explanations\\
				\hline
				$\mathcal{N},n$ & road network, number of vertices of the road network \\
				$G$ & the adjacency matrix of a graph \\
				$U$ & latent space matrix \\
				$B$ & attribute interaction matrix\\
				$A$ & the transition matrix\\			
				$k$ & the number of dimensions of latent attributes\\
				$T$ & the number of snapshots\\
				$span$ & the gap between two continuous graph snapshots\\
				$h$ & the prediction horizon\\
				$\lambda, \gamma$ & regularization parameters for graph Laplacian and transition process\\
				\hline	
			\end{tabular}}
		\end{small}
		\vspace{-5mm}
	\end{table}

\vspace{-2mm}
\section{Problem definition}\label{sec-problem}

We denote a road network as a directed graph $\mathcal{N}=(V,E)$, where $V$ is the set of vertices and $E\in V \times V$ is the set of edges, respectively. A vertex $v_i\in V$ models a road intersection or an end of road. An edge $e(v_i, v_j)$, which connects two vertices, represents a directed network segment. Each edge $e(v_i, v_j)$ is associated with a travel speed $c(v_i, v_j)$ (e.g., 40 miles/hour). In addition, $\mathcal{N}$ has a corresponding adjacency matrix representation, denoted as $G$, whose $(i,j)^{\texttt{th}}$ entry represents the edge weight between the $i^{\texttt{th}}$ and $j^{\texttt{th}}$ vertices.

The road network snapshots are constructed from a large-scale, high resolution traffic sensor dataset (see detailed description of sensor data in Section~\ref{sec-exp}). Specifically, a sensor $s$ (i.e., a loop detector) is located at one segment of road network $\mathcal{N}$, which provides a reading (e.g., 40 miles/hour) per sampling rate (e.g., 1 min). We divide one day into different intervals, where $span$ is the length of each time interval. For example, when $span=5$ minutes, we have 288 time intervals per day. For each time interval $t$, we aggregate the readings of one sensor. Subsequently, for each edge segment of network $\mathcal{N}$, we aggregate all sensor readings located at that edge as its weight. Therefore, at each timestamp $t$, we have a road network snapshot $G_t$ from traffic sensors. 

\begin{figure}[!ht]
	\begin{small}\scalebox{0.9}{
			\begin{tabular}{cc}
				\vspace{-0mm}\hspace{-0mm}\includegraphics[height=18mm]{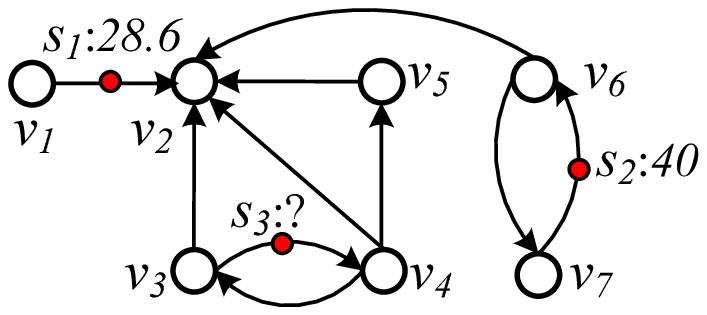}&
				\vspace{+2mm}\hspace{+6mm}\includegraphics[height=20mm]{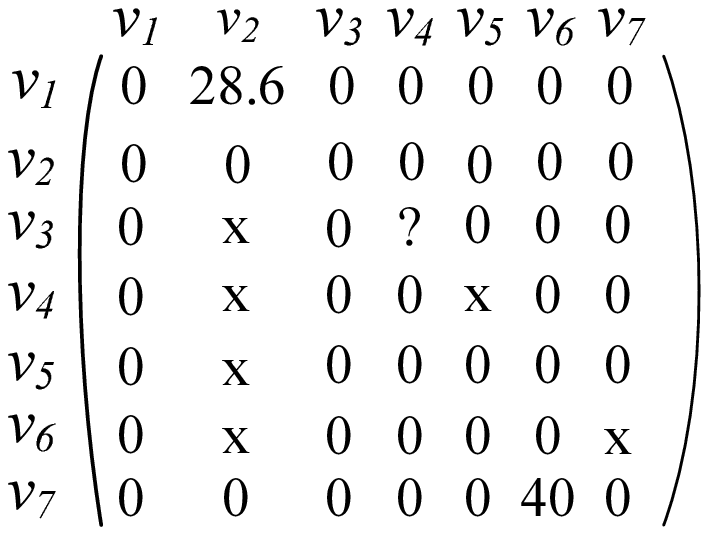} \\
				\hspace{-3mm}(a) An abstract road network $\mathcal{N}$ & \hspace{-0mm}(b) Adjacency matrix representation $G$
			\end{tabular}}
		\end{small}
		\vspace{-3mm}\caption{\small An example of road network} \vspace{-0mm}
		\label{fig::rn}
		\vspace{-4mm}
\end{figure}

\vspace{+1mm}
\noindent\textbf{Example.} \textit{Figure~\ref{fig::rn} (a) shows a simple road network with 7 vertices and 10 edges at one timestamp. Three sensors (i.e., $s_1, s_2, s_3$) are located in edges $(v_1, v_2)$, $(v_3, v_4)$ and $(v_7, v_6)$ respectively, and each sensor provides an aggregated reading during the time interval. Figure~\ref{fig::rn}(b) shows the corresponding adjacent matrix after mapping the sensor readings to the road segments. Note that the sensor dataset is incomplete with both missing values (i.e., sensor fails to report data) and missing sensors (i.e., edges without any sensors). Here sensor $s_3$ fails to provide reading, thus the edge weight of $c(v_3, v_4)$ is $?$ due to missing value. In addition, the edge weight of $c(v_3, v_2)$ is marked as $\times$ because of missing sensors.}

\vspace{-0mm}

Given a small number of road network snapshots, or a dynamic road network, our objective is to predict the future traffic conditions. Specifically, a \textit{dynamic road network}, is a sequence of snapshots $(G_1, G_2, \cdots, G_T)$ with edge weights denoting time-dependent travel cost.

With a dynamic road network, we formally define the problem of edge traffic prediction with missing data as follows:
\vspace{-1mm}
\begin{problem}
	Given a dynamic road network $(G_1, G_2, \cdots, G_T)$ with missing data at each timestamp,
	we aim to achieve the following two goals:
	\begin{itemize}
		\vspace{-1mm}
		\item complete the missing data (i.e., both missing value and sensor) of $G_i$ , where $1 \leq i \leq T$;
		\vspace{-1mm}
		\item predict the future readings of $G_{T+h}$, where $h$ is the prediction horizon. For example, when $h=1$, we predict the traffic condition of $G_{T+1}$ at the next timestamp.
	\end{itemize}
\end{problem}

For ease of presentation, Table~\ref{tbl:notation} lists the notations we use throughout this paper. Note that since each dimension of a latent space represents a latent attribute, we thus use latent attributes and latent positions interchangeably. 

\vspace{-3mm}
\section{Latent Space Model for Road Networks (LSM-RN)}\label{sec-model}
In this section, we describe our LSM-RN model in the context of traffic prediction. We first introduce the basic latent space model (Section~\ref{sub:spatialmodel}) by considering the graph topology, and then incorporate both temporal and transition patterns (Section~\ref{subsec:temporal}).  Finally, we describe the complete LSM-RN model to solve the traffic prediction problem with missing data (Section~\ref{subsec:predict}). 

\vspace{-1mm}
\subsection{Topology in LSM-RN }\label{sub:spatialmodel}	

Our traffic model is built upon the latent space model of the observed road network. Basically, each vertex of road network have different attributes and each vertex has an overlapping representation of attributes. The attributes of vertices and how each attribute interacts with others jointly determine the underlying traffic patterns. Intuitively, if two highway vertices are connected, their corresponding interaction generates a higher travel speed than that of two vertices located at arterial streets. In particular, given a snapshot of road network $G$, we aim to learn two matrices $U$ and $B$, where matrix $U \in R_{+}^{n\times k}$ denotes the latent attributes of vertices, and matrix $B \in R_{+}^{k \times k}$ denotes the attribute interaction patterns. The product of $UBU^T$ represents the traffic speed between any two vertices, where we use to approximate $G$. Note that $B$ is an asymmetric matrix since the road network $G$ is directed. Therefore, the basic traffic model which considers the graph topology can be determined by solving the following optimization problem:

\vspace{-2mm}
\begin{small}
	\begin{equation}
	\argmin_{U\geq 0,B\geq 0}J=||G-UBU^T||_F^2
	\end{equation}
\end{small}
\vspace{-2mm}

\begin{figure}[!ht]
	\begin{small}\scalebox{0.90}{
			\begin{tabular}{cc}
				\hspace{-1mm}\includegraphics[width=40mm]{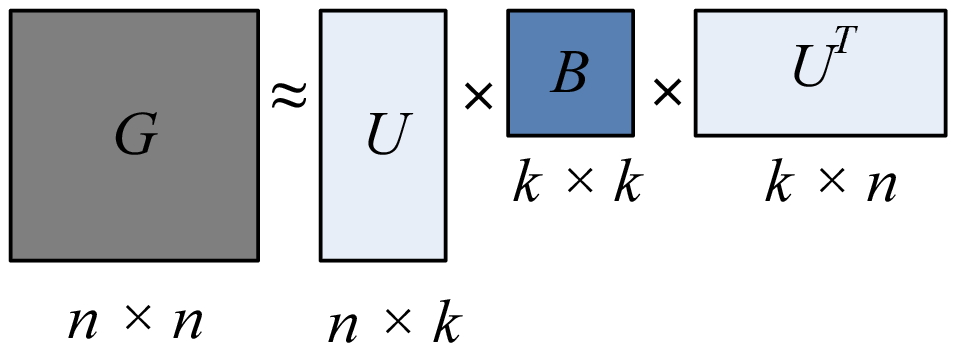} &
				\vspace{-1mm}\hspace{+2mm}\includegraphics[width=40mm]{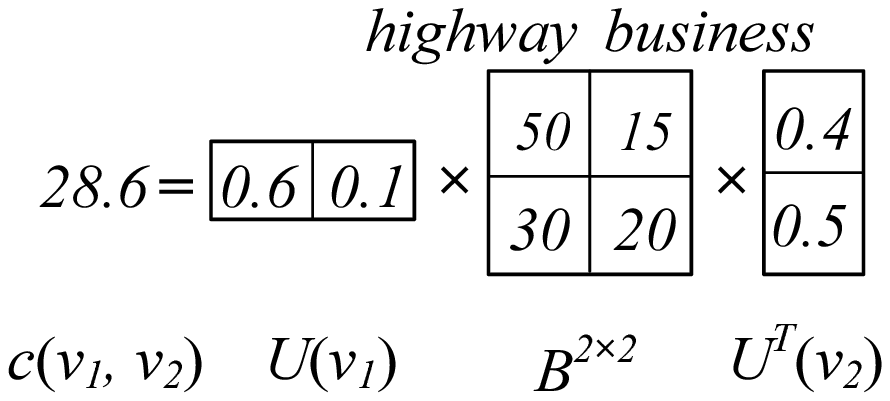}  \vspace{1mm}\\
				\hspace{-6mm}(a) Basic model & \hspace{+3mm}(b) Travel time of $c(v_1, v_2)$
			\end{tabular}}
		\end{small}
		\vspace{-2mm}\caption{\small An example of our traffic model, where $G$ represents a road network, $U$ denotes the attributes of vertices in the road network, $n$ is number of nodes, and $k$ is number of attributes, and $B$ denotes how one type of attributes interacts with others. } \vspace{-0mm}
		\label{fig::model-example}
		\vspace{-2mm}
	\end{figure}\vspace{-1mm}
	
	Figure~\ref{fig::model-example} (a) illustrates the intuition of our static traffic model. 
	As shown in Figure~\ref{fig::model-example} (b), suppose we know that each vertex is associated with two attributes (e.g., highway and business area), and the interaction pattern between two attributes is encoded in matrix $B$, we can accurately estimate the travel speed between vertex $v_1$ and $v_2$, using their corresponding latent attributes and the matrix $B$.
	
	\vspace{0.1cm}
	\noindent\textbf{Overcome the sparsity of $G$.} In our road network, $G$ is very sparse (i.e., zero entries dominate the items in $G$) 
	for the following reasons: (1) the average degree of a road network is small~\cite{wu2012shortest}, and thus the edges of road network is far from fully connected,
	(2) the distribution of sensors is non-uniform, and only a small number of edges are equipped with sensors; and (3) there exists missing values (for those edges equipped with sensors) due to the failure and/or maintenance of sensors.
	
	
	
	
	Therefore, we define our loss function only on edges with observed readings, that is, the set of edges with travel cost $c(v_i, v_j)>0$. In addition, we also propose an in-filling method to reduce the gap between the input road network and the estimated road network. 
	We consider 
	graph Laplacian dynamics, which is an effective smoothing
	approach for finding global structure similarity~\cite{lambiotte2008laplacian}. Specifically, we construct a graph Laplacian matrix $L$, defined as $L=D-W$, where $W$ is a graph proximity matrix that is constructed from the network topology, and $D$ is a diagonal matrix $D_{ii}=\sum_{j}(W_{ij})$. With these new constraints, our traffic model for one snapshot of road network $G$ is expressed as follows:
	
	\vspace{-1mm}
	\begin{small}
		\begin{equation}
		\argmin_{U,B}J=||Y\odot (G-UBU^T)||_F^2 + \lambda Tr(U^TLU)
		\end{equation}
	\end{small}
	\vspace{-1mm}
	
	where $Y$ is an indication matrix for all the non-zero entries in $G$, i.e, $Y_{ij} = 1$ if and only if $G(i,j) > 0$; $\odot$ is the Hadamard product operator, i.e., $(X\odot Z)_{ij}=X_{ij} \times Z_{ij}$; and $\lambda$ is the Laplacian regularization parameter.

	\subsection{Time in LSM-RN}\label{subsec:temporal}
	We now combine the temporal information, including time-dependent modeling of latent attributes and the temporal transition. With this model, each vertex is represented in a unified latent space, where each dimension either represents a spatial or temporal attribute.
	
	
	\vspace{-1mm}
	\subsubsection{Temporal effect of latent attributes}
	The behavior of the vertices of road networks may evolve pretty quickly. For instance, the behavior of a vertex that is similar to that of a highway vertex during normal traffic condition, may become similar to that of an arterial street node during congestion hours. Because the behavior of each vertex can change over time,  we must  employ a time-dependent modeling for attributes of vertices for real-time traffic prediction. Therefore, we add the time-dependent effect of attributes into our traffic model. Specifically, for each $t \leq T$, we aim to learn a corresponding time-dependent latent attribute representation $U_t$. Although the latent attribute matrix $U_t$ is time-dependent, we assume that the attribute interaction matrix $B$ is an inherent property, and thus we opt to fix $B$ for all timestamps. 
	By incorporating this temporal effect, we obtain our model based on the following optimization problem:
	
	\vspace{-4mm}
	\begin{small}
		\begin{equation}
		\begin{aligned}
		\argmin_{U_t,B}J=&\sum_{t=1}^{T}||Y_t\odot (G_t-U_tBU_t^T)||_F^2 +\sum_{t=1}^{T}\lambda Tr(U_tLU_t^T)
		\end{aligned}	
		\end{equation}
		\vspace{-2mm}
	\end{small}

	\vspace{-5mm}
	\subsubsection{Transition matrix}
	Due to the dynamics of traffic condition, we not only want to learn the time-dependent latent attributes, but also learn a transition model to capture the evolving behavior from one snapshot to the next. The transition should be able to capture both periodic evolving patterns (e.g., morning/afternoon rush hours) and non-recurring patterns caused by traffic incidents (e.g., accidents, road construction, or work zone closures). For example, during the interval of an accident, a vertex transition from normal state to congested at the beginning, then become normal again after the accident is cleared.
	
	We thus assume a global  process to capture the state transitions. Specifically, we use a matrix $A$ that approximates the changes of $U$ between time $t-1$ to time $t$, i.e., $U_{t} = U_{t-1} A$, where $U \in R_+^{n\times k}, A \in R_+^{k \times k}$. The transition matrix $A$ represents how likely a vertex is to transit from attribute $i$ to attribute $j$ for that particular time interval.
	
	\subsection{LSM-RN Model}  \label{subsec:predict}
	
	Considering all the above discussions, the final objective function for our LSM-RN model is defined as follows:
	
	\vspace{-3mm}
	\begin{small}
		\begin{equation}
		\label{eqn::final-obj}
		\begin{aligned}
		\argmin_{U_t,B,A}J=&\sum_{t=1}^{T}||Y_t\odot (G_t-U_tBU_t^T)||_F^2 + \sum_{t=1}^{T}\lambda Tr(U_tLU_t^T) +\\ &\sum_{t=2}^T\gamma||U_t-U_{t-1}A||_F^2
		\end{aligned}
		\end{equation}
		\vspace{-1mm}
	\end{small}
	\vspace{-4mm}
	
	where $\lambda$ and $\gamma$ are the regularization parameters.
	
	\subsubsection{Edge traffic prediction with missing data}

	Suppose by solving Equation~\ref{eqn::final-obj}, we obtain the learned matrices of $U_t, B$ and $A$ from our LSM-RN model. Consequently, the task of both missing value and sensor completion can be accomplished by the following:
	
	\vspace{-1mm}
	\begin{small}
		\begin{equation}
		\begin{aligned}
		G_t=&U_tBU_t^T, \quad\quad \mbox{when $1\leq t \leq T$}
		\end{aligned}
		\end{equation}
		\vspace{-2mm}
	\end{small}

	Subsequently, the edge traffic for snapshot $G_{T+h}$, where $h$ is the number of future time spans, can be predicted as follows:
	\begin{small}
		\begin{equation}
		\begin{aligned}
		G_{T+h}=&(U_TA^h)B(U_TA^h)^T 
		\end{aligned}
		\end{equation} 
		\vspace{-2mm}
	\end{small}

\begin{figure}[!ht]
	\centering
	\includegraphics[width=70mm]{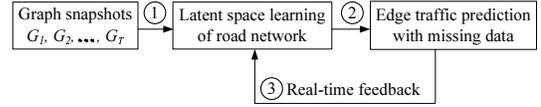}
	\vspace{-3mm}\caption{Overall Framework} \vspace{-2mm} \label{fig::framework}
\end{figure}
Figure~\ref{fig::framework} shows an overview of the LSM-RN framework. Given a series of road network snapshots, LSM-RN processes them in three steps:  (1) discover the latent attributes of vertices at each timestamp, which capture both the spatial and temporal properties; (2) understand the traffic patterns and build a predictive model of how these latent attributes change over time; and (3) exploit real time traffic information to adjust the existing latent space models.

\section{Inference of LSM-RN}\label{sec-global}
In this section, we first propose a global multiplicative algorithm, and then discuss a fast incremental algorithm that scales to large road networks.

\subsection{Global Learning Algorithm}
We develop an iterative update algorithm to solve Equation~\ref{eqn::final-obj}, which belongs to the category of traditional multiplicative update algorithm~\cite{lee2001algorithms}. 
By adopting the methods from~\cite{lee2001algorithms}, we can derive the update rule of $U_t$, $B$ and $A$.

%
%
%

\subsubsection{Update rule of $U_t$}
We first consider updating variable $U_t$ while fixing all the other variables. The part of objective function in Equation~\ref{eqn::final-obj} that is related to $U_t$ can be rewritten as follows:

\vspace{-3mm}
\scalebox{0.88}{
	\begin{minipage}{.\linewidth}
		\begin{equation*}
		\small{
			\begin{aligned}
			&J = \sum_{t=1}^{T}Tr\Big(\big(Y_t\odot(G_t-U_tBU_t^T)\big)\big(Y_t\odot(G_t-U_tBU_t^T)\big)^T\Big) \\
			&+\sum_{t=1}^{T}\lambda Tr(U_t(D-W)U_t^T)+\sum_{t=2}^{T}Tr\Big(\gamma(U_t-U_{t-1}A)(U_t-U_{t-1}A)^T\Big)
			\end{aligned}
		}
		\end{equation*}
	\end{minipage}
}

Because we have the non-negative constraint of $U_t$, following the standard constrained optimization theory, we introduce the Lagrangian multiplier $(\psi_t) \in R^{n\times k}$ and minimize the Lagrangian function $L$:

\begin{small}
\begin{equation}\label{eqn::obj-L}
\begin{aligned}
L = J + \sum_{t=1}^{T}Tr(\psi_tU_t^T)
\end{aligned}
\end{equation}
\end{small}

Take the derivation of $L$ with respect to $U_t$, we have the following expression. (The detail is described in Appendix~\ref{sec::app-1})

\begin{small}
\hspace{-5mm}
\begin{equation}
\hspace{-3mm}
\begin{aligned}
\hspace{-5mm}&\frac{\partial L}{\partial U_t}=-2(Y_t\odot G_t)U_t B^T-2(Y_t^T\odot G_t^T)U_t B+2(Y_t\odot U_tBU_t^T)(U_tB^T+U_tB)\\
\hspace{-5mm}&+2\lambda (D-W)U_{t}+ 2\gamma(U_t-U_{t-1}A) + 2\gamma(U_{t}AA^T-U_{t+1}A^T)+\psi_{t} \\
\end{aligned}
\end{equation}
\end{small}

By setting $\frac{\partial L}{\partial U_t}=0$, and using the KKT conditions $(\psi_t)_{ij}(U_t)_{ij}=0$,
we obtain the following equations for $(U_t)_{ij}$:

\vspace{-2mm}
\begin{small}
\begin{equation}
\begin{aligned}
&\big[-(Y_t\odot G_t)U_t B^T-(Y_t^T\odot G_t^T)U_t B)+(Y_t\odot U_tBU_t^T)(U_tB^T+U_tB)\\
&+\lambda LU_t+ \gamma(U_\tau-U_{\tau-1}A) + \gamma(U_{\tau}AA^T-U_{\tau+1}A^T)\big]_{ij}(U_t)_{ij} = 0 
\end{aligned}
\end{equation}
\end{small}

Following the updating rules proposed and proved in~\cite{lee2001algorithms}, we
derive the following update rule of $U_t$:

\scalebox{0.88}{
\hspace{-8mm}
\begin{minipage}{.\linewidth}
\begin{equation}\label{eqn::up_u}
\begin{aligned}
\hspace{-4mm}(U_{t})\leftarrow & (U_{t})\odot\\
& \Big(\frac{(Y_t\odot G_t)U_{t}B^T+(Y_t^T\odot G_t^T)U_{t}B+\lambda WU_t+\gamma (U_{t-1}A+U_{t+1}A^T)}{(Y_t\odot U_{t}BU_{t}^T)(U_{t}B^T+U_{t}B)+\lambda DU_t+\gamma(U_{t}+U_{t}AA^T)}\Big)^\frac{1}{4}\\
\end{aligned}
\end{equation}
\end{minipage}
}

\subsubsection{Update rule of $B$ and $A$}
The updating rules for $A$ and $B$ could be derived as follows in a similar way (see Appendix~\ref{sec::app-2} for detailed calculation):

\begin{equation}\label{eqn::up_b}
\begin{aligned}
&B\leftarrow B \odot \Big(\frac{\sum_{t=1}^{T}U_{t}^T(Y_t\odot G_{t}) U_{t}}{\sum_{t=1}^{T}U_{t}^T(Y_t\odot (U_{t}BU_{t}^T))U_{t}}\Big)\\
\end{aligned}
\end{equation}

\begin{equation}\label{eqn::up_a}
\begin{aligned}	
&A\leftarrow A \odot \Big(\frac{\sum_{t=1}^{T}U_{t-1}^TU_{t}}{\sum_{t=1}^{T}U_{t-1}^TU_{t-1}A}\Big)\\
\end{aligned}
\end{equation}

\begin{algorithm}[!t]\small\caption{Global-learning($G_1, G_2, \cdots, G_T$)}
	\textbf{Input:} graph matrix $G_1, G_2, \cdots, G_T$.\\
	\textbf{Output:}  $U_t$ ($1\leq t \leq T$), $A$ and $B$. \\
	\vspace{-2mm}
	\label{alg::temporal-mu}
	\begin{algorithmic}[1]
		\STATE{Initialize $U_{t}$, $B$ and $A$}
		\WHILE{Not Convergent}
			\FOR{$t = 1$ to $T$}
				\STATE{update $U_t$ according to equation~\ref{eqn::up_u}}
			\ENDFOR
			\STATE{update $B$ according to Equation~\ref{eqn::up_b}}
			\STATE{update $A$ according to Equation~\ref{eqn::up_a}}
		\ENDWHILE
	\end{algorithmic}
\end{algorithm}

Algorithm~\ref{alg::temporal-mu} outlines the process of updating each matrix using aforementioned multiplicative rules to optimize Eq.~\ref{eqn::final-obj}. The general idea is to jointly infer and cyclically update all the latent attribute matrices $U_t$, $B$ and $A$. In particular, we first jointly learn the latent attributes for each time $t$ from all the graph snapshots (Lines  2--4). Based on the sequence of time-dependent latent attributes $<U_1, U_2, \cdots, U_T>$, we then learn the global attribute interaction pattern $B$ and the transition matrix $A$ (Lines 6--7).

From Algorithm~\ref{alg::temporal-mu}, we now explain how our LSM-RN model jointly learns the spatial and temporal properties. Specifically, when we update the latent attribute of one vertex $U_t{(i)}$, the spatial property is preserved by (1) considering the latent positions of its adjacent vertices ($Y_t\odot G_{t}$), and (2) incorporating the local graph Laplacian constraint (i.e., matrix $W$ and $D$). Moreover, the temporal property of one vertex is then captured by leveraging its latent attribute in the previous and next timestamps (i.e.,  $U_{t-1}(i)$ and $U_{t+1}(i)$), as well as the transition matrix.

In the following, we briefly discuss the time complexity adn convergence for our global multiplicative algorithm. In each iteration,  the computation is dominated by matrix multiplication operations:  the matrix multiplication between a $n\times n$ matrix and a $n\times k$ matrix, and another matrix multiplicative operator between a $n\times k$ matrix and a $k\times k$ matrix. Therefore, the worst case time complexity per iteration is dominated by $O(T(nk^2+n^2k))$. However, since all the matrices are sparse, the complexity of matrix multiplication with two $n\times k$ sparse matrix, is much smaller than $O(n^2k)$. Followed the proof shown in previous works~\cite{CaiPAMI2011}~\cite{lee2001algorithms}~\cite{zhu2014tripartite}, we could prove that  Algorithm~\ref{alg::temporal-mu} converges into a local minimal and the objective value is non-increasing in each iteration.

\begin{figure}[!ht]
	\begin{small}\scalebox{0.99}{
			\begin{tabular}{cc}
				\hspace{-1mm}\includegraphics[width=42mm]{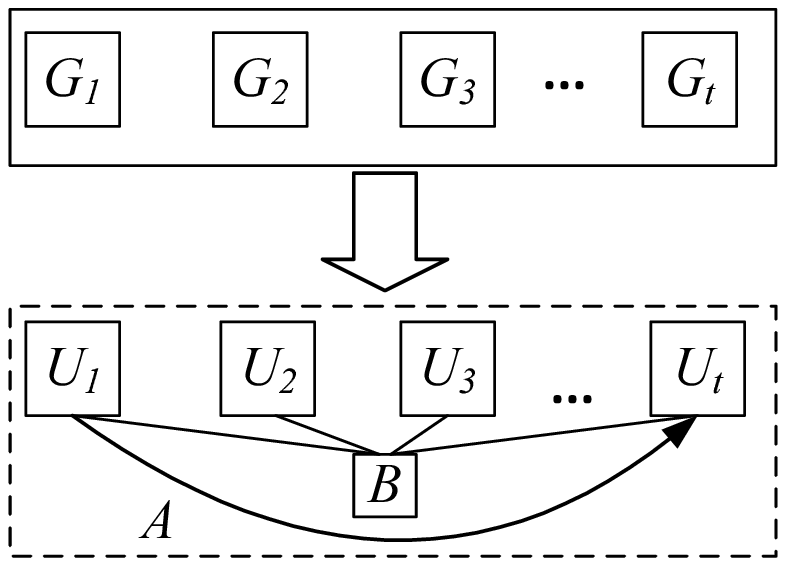} &
				\vspace{-1mm}\hspace{+2mm}\includegraphics[width=42mm]{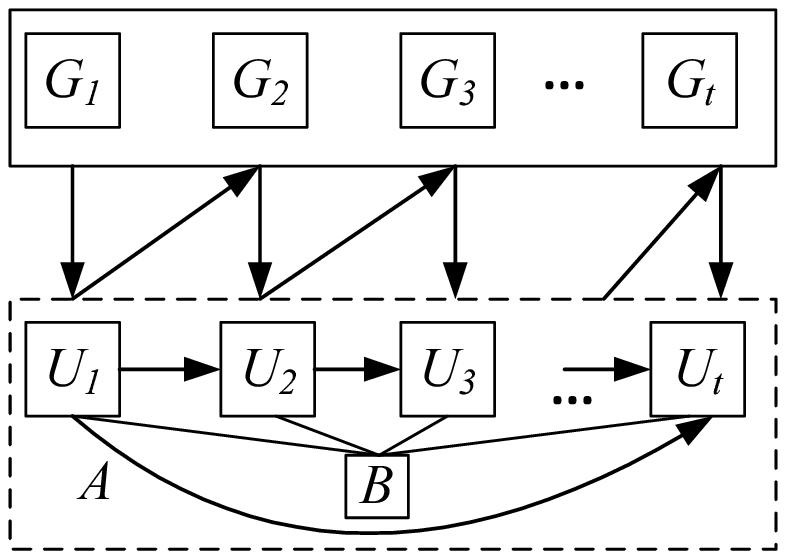}  \vspace{1mm}\\
				\hspace{-6mm}(a) Global learning & \hspace{-5mm}(b) Incremental learning
			\end{tabular}}
		\end{small}
		\vspace{-2mm}\caption{\small Illustration of algorithms} \vspace{-0mm}
		\label{fig::global-incremental}
		\vspace{-0mm}
\end{figure}\vspace{-0mm}

\subsection{Incremental learning algorithm}\label{sec-incremental}
Although global multiplicative algorithm accurately captures the latent attribute, it is computationally expensive. As illustrated in Figure~\ref{fig::global-incremental} (a), the latent attributes are jointly inferred from the entire set of graph snapshots and cyclically updated until they become stable. Unfortunately, this joint and cyclic inference is very time consuming. A naive method is to treat each snapshot independently, and learn the temporal latent attribute at each timestamp separately without considering the temporal relationships.
To improve the efficiency as well as preserve the topology and temporal properties, we propose an incremental algorithm. As depicted in Figure~\ref{fig::global-incremental} (b), our incremental algorithm  sequentially and adaptively learns the latent attributes from the temporal graph changes.

\subsubsection{Framework of incremental algorithm}
The intuition of our incremental algorithm is based on the following observation: during a short time interval (e.g., 5 minutes), the overall traffic condition of the whole network tends to stay steady, and the travel costs of most edges change at a slow pace. For those edges with minor travel speed variations, their corresponding positions in the latent space do not change much either. Nevertheless, we still need to identify vertices with obvious variations in terms of their travel speeds, and adjust their corresponding latent attributes. For example, some edges could be in a transition state from non-rush hour to rush hour, and thus the travel speed reduces significantly. Therefore, instead of recomputing the latent attribute of each vertex from scratch at every time stamp, we perform ``lazy" adjustment, utilizing the latent attributes we have already learned in the previous snapshot. As shown in Figure~\ref{fig::global-incremental}, to learn the latent attribute of $U_t$, the incremental algorithm utilizes the latent attributes we already learned in the previous snapshot (i.e., $U_{t-1}$) and the dynamism of traffic condition.



Algorithm~\ref{alg::incremental-mu} presents the pseudo-code of incremental learning algorithm. Initially, we learn the latent space of $U_1$ from our global multiplicative algorithm (Line 1). With the learned latent matrix $U_{t-1}$, at each time stamp $t$ between $2$ and $T$, we incrementally update the latent space of $U_t$ from $U_{t-1}$ according to the observed graph changes (Lines 2-4). After that, we learn the global transition matrix $A$ (Line 5). 

\begin{algorithm}[!ht]\small\caption{Incremental-Learning($G_1, G_2, \cdots, G_T$)}
	\textbf{Input:} graph matrix $G_1, G_2, \cdots, G_T$.\\
	\textbf{Output:}  $U_t$ ($1\leq t \leq T$), $A$ and $B$. \\
	\vspace{-2mm}
	\label{alg::incremental-mu}
	\begin{algorithmic}[1]
		\STATE{$(U_{1}, B) \leftarrow$Global-learning($G_1$)} 
		\FOR{$t = 2$ to $T$}
			\STATE{$U_t \leftarrow$ Incremental-Update($U_{t-1}, G_t$) (See Section~\ref{sec::inc-update})}
		\ENDFOR
		\STATE{Iteratively learn transition matrix $A$ using Equation~\ref{eqn::up_a} until $A$ converges}
	\end{algorithmic}
\end{algorithm}

%

\subsubsection{Topology-aware incremental update} \label{sec::inc-update}
Given $U_{t-1}$ and $G_{t}$, we now explain how to calculate $U_{t}$ incrementally from $U_{t-1}$ 
, with which we could accurately approximate $G_{t}$. The main idea is similar to an online learning process. 
At each round, the algorithm predicts an outcome for the required task (i.e., predict the speed of edges). Once the algorithm makes a prediction, it receives feedback indicating the correct outcome. Then, the online algorithm can modify its prediction mechanism, presumably improving the changes of making a more accurate prediction on subsequent timestamps. In our scenario, we first use the latent attribute matrix $U_{t-1}$ to make a prediction of $G_t$ as if we do not know the observation, subsequently we adjust the model of $U_t$ according to the true observation of $G_t$ we already have in hand.

However, in our problem, we are making predictions for the entire road network, not for a single edge. When we predict for one edge, we only need to adjust the latent attributes of two vertices, whereas in our scenario we need to update the latent attributes for many correlated vertices. Therefore, the effect of adjusting the latent attribute of one vertex could potentially affects its neighboring vertices, and influences the convergence speed of the incremental algorithm. Hence, the adjustment order of vertices also matters in our incremental update.

In a nutshell, our incremental update consists of the following two components: 1) identify candidate nodes based on feedbacks; 2) update their latent attributes and propagate the adjustment from one vertex to its neighbors. As outlined in Algorithm~\ref{alg::incremental-update}, given $U_{t-1}$ and $G_t$, we first make an estimation of $\widehat{G_t}$ based on $U_{t-1}$ (Line 1),
subsequently we treat $G_t$ as the feedback information, and select the set of vertices where we make inaccurate 
predictions, and insert them into a candidate set $cand$ (Lines 3-9). Consequently, for each vertex $i$ of $cand$, we adjust its latent attribute so that we could make more accurate predictions (Line 15) and then examine how this adjustment would influence the candidate task set from the following two aspects: (1) if the latent attribute of $i$ does not change much, we remove it from the set of $cand$ (Lines 17-19); (2) if the adjustment of $i$ also affects its neighbor $j$, we add vertex $j$ to $cand$ (Lines 20-25).

\begin{algorithm}[!ht]\small\caption{Incremental-Update($U_{t-1}, G_{t}$)}
	\textbf{Input:} the latent matrix $U_{t-1}$ at $t-1$, Observed graph reading $G_{t}$\\
	\textbf{Output:} Updated latent space $U_{t}$. \\
	\vspace{-2mm}
	\label{alg::incremental-update}
	\begin{algorithmic}[1]
		\STATE{$\widehat{G_t} \leftarrow U_{t-1}BU_{t-1}^T$}
		\STATE{$cand \leftarrow \emptyset$}
		\FOR{each $i \in G$}
		\FOR{each $j \in out(i)$}
		\IF{$|G_t(i,j) -\widehat{G_t(i,j)}| \ge \delta $}
		\STATE{$cand \leftarrow cand \cup \{i,j\}$}
		\ENDIF
		\ENDFOR
		\ENDFOR
		\STATE{$U_{t} \leftarrow U_{t-1}$}
		\WHILE{Not Convergent AND $cand \notin \emptyset$}
		\FOR{$i \in cand$} 
		\STATE{$oldu \leftarrow U_t(i)$}
		\FOR{each $j \in out(i)$}
		\STATE{adjust $U_{t}(i)$  with Eq.~\ref{eq:increonline}}
		\ENDFOR
		\IF{$||U_{t}(i)-oldu||_F^2 \leq \phi$}
		\STATE{$cand \leftarrow cand \setminus \{i\}$}
		\ENDIF
		\FOR{each $j \in out(i)$}
		\STATE{$p \leftarrow U_{t}(i)BU_{t}(j)$}
		\IF{$|p-{G_t(i,j)}|\geq \delta $}
		\STATE{$cand \leftarrow cand \cup \{j\}$}
		\ENDIF
		\ENDFOR
		\ENDFOR             
		\ENDWHILE
	\end{algorithmic}
\end{algorithm}

The remaining questions in our Incremental-Update algorithm are how to adjust the latent position of one vertex according to feedbacks, and how to decide the order of update. In the following, we address each of them separately.

\begin{figure}[!ht]
	\begin{small}\scalebox{0.99}{
			\begin{tabular}{cc}
				\hspace{-6mm}\includegraphics[width=38mm]{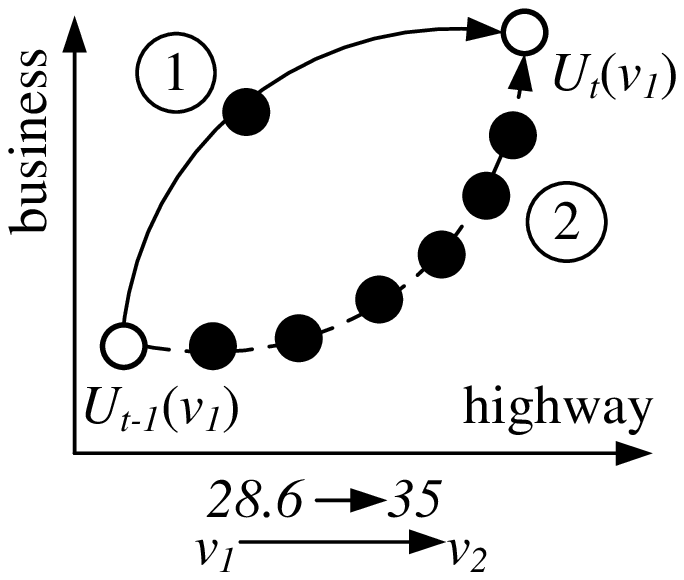} &
				\vspace{-1mm}\hspace{+6mm}\includegraphics[width=38mm]{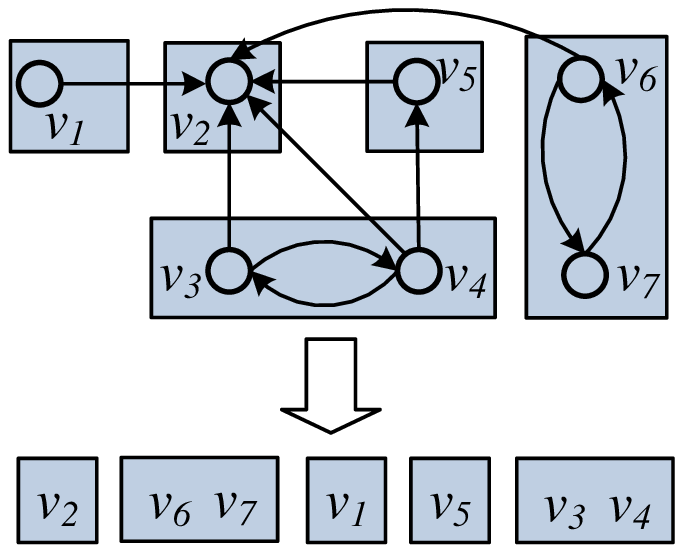}  \vspace{1mm}\\
				\hspace{-6mm}(a) Adjustment method & \hspace{-5mm}(b) Adjustment order
			\end{tabular}}
		\end{small}
		\vspace{-2mm}\caption{\small Two challenges of adjusting the latent attribute with feedbacks.} \vspace{-0mm}
		\label{fig::incremental-example}
		\vspace{-0mm}
	\end{figure}\vspace{-2mm}

\vspace{+2mm}		
\noindent\textbf{Adjust the latent attribute of one vertex.}
To achieve high efficiency of adjusting the latent attribute, we propose to make the smallest changes of the latent space (as quick as possible) to predict the correct value. For example, as shown in Figure~\ref{fig::incremental-example}(a), suppose we already know the new latent position of $v_1$, then fewer step movement (Option 1) is preferable than gradual adjustment (Option 2). Note that in our problem, when we move the latent position of a vertex to a new position, the objective of this movement is to produce a correct prediction for each of its outgoing edges. Specifically, given $U_{t-1}(i)$, we want to find $U_t(i)$ which could accurately predict the weight of each edge $e(v_i,v_j)$ that is adjacent to vertex $v_i$. We thus formulate our problem as follows:



\begin{equation}
\begin{aligned}
U_{t}(i),\xi^* &=\arg\min_{U(i)\in R_+^k}\frac{1}{2}||U(i)-U_{t-1}(i)||_F^2 + C\xi\\
\mbox{s.t.} \quad &|U(i)BU^T(j)-G_t(i,j)| \leq \delta+\xi
\end{aligned}
\end{equation}

Note that we have non-negativity constraint over the latent space of $U_{t}(i)$. To avoid over aggressive update of the latent space, we add a non-negative slack variable $\xi$ into the optimization problem.  Therefore, we adopt the approaches from~\cite{blondel2014online}: When the predicted value $\widehat{y_t}$ (i.e., $U_t(i)BU_t^T(j)$) is less than the correct value $y_t$ (i.e., $G_t(i,j)$), we use the traditional online passive-aggressive algorithm~\cite{crammer2006online} because it still guarantees the non-negativity of $U(i)$; Otherwise, we update $U(i)$ by solving a quadratic optimization problem. The detailed solution is as follows:

\begin{equation}\label{eq:increonline}
\begin{aligned}
U_{t}(i)=\max(U_{t-1}(i)+(k^*-\theta^*)\cdot BU_{t-1}(j)^T, 0)
\end{aligned}
\end{equation}

$k^*$ and $\theta^*$ are computed as follows:
\begin{equation}
\left\{
\begin{tabular}{ll}
$k^*=\alpha_t, \theta^*=0$ & if $\widehat{y_t} < y_t$  \\
$k^*=0, \theta^*=C$ & if $\widehat{y_t} > y_t$ and $f(C)\geq 0$  \\
$k^*=0, \theta^*=f^{-1}(0)$ & if $\widehat{y_t} > y_t$ and $f(C)< 0$  
\end{tabular}
\right.
\end{equation}

where 
\begin{equation*}
\begin{aligned}
\alpha_t&=\min\Big(C,\frac{\max(|\widehat{y_t}-y_t|-\delta,0)}{||BU_{t-1}(j)^T||^2}\Big)\\
f_t(\theta)&=\max\big(U_t(i)-\theta BU_{t}(j)^T, 0\big)\cdot BU_{t}(j)^T -G_t(i,j)-\delta
\end{aligned}
\end{equation*}

\noindent\textbf{Updating order of $cand$.}
As we already discussed, the update order is very important because it influences the convergence speed of our incremental algorithm. Take the example of the road network shown in Figure~\ref{fig::rn}, suppose our initial $cand$ contains three vertices $v_7, v_6$ and $v_2$, where we have two edges $e(v_7,v_6)$ and $e(v_6,v_2)$. If we randomly choose the update sequence as $<v_7,v_6,v_2>$, that is, we first adjust the latent attribute of $v_7$ so that $c(v_7,v_6)$ has a correct reading; subsequently we adjust the latent attribute of $v_6$ to correct our estimation of $c(v_6,v_2)$. Unfortunately,the adjustment of $v_6$ could influence the correction we have already made to $v_7$, thus leading to an inaccurate estimation of $c(v_7,v_6)$ again. A desirable order is to first update vertex $v_6$ before updating $v_7$.


Therefore, we propose to consider the reverse topology of the road network when we update the latent position of each candidate vertex $v\in cand$. The general principle is that: given edge $e(v_i,v_j)$, the update of vertex $v_i$ should be proceeded after the update of $v_j$, since the position of $v_i$ is dependent on  $v_j$.
This motivates us to derive a reverse topological order in the graph of $G$. Unfortunately, the road network $G$ is not a Directed Acyclic Graph (DAG), and contains cycles. To address this issue, we first generate Strongly Connected Components (SCC) of the graph $G$, and contract each SCC as a super node. We then derive a topological order based on this condensed graph. For the vertex order in each SCC, we can either decide randomly, or use some heuristics to break the SCC into a DAG, thus generating an ordering of vertices inside each SCC. Figure~\ref{fig::incremental-example}(b) shows an example of ordering for the road network of Figure~\ref{fig::rn}, where each rectangle represents a SCC. After generating a reverse topological order based on the contracted graph and randomly decide the vertex order of each SCC, we obtain one final ordering $<v_2,v_6,v_7,v_1,v_5,v_4,v_3>$. Therefore, each time when we update the latent attributes of $cand$, we follow this ordering of vertices.



%
%

\vspace{+2mm}
\noindent\textbf{Time complexity} For each vertex $i$, the computational complexity of adjusting its latent attributes using Eq.~\ref{eq:increonline} is $O(k)$, where $k$ is number of attributes. Therefore, to compute latent attributes $u$, the time complexity per iteration is $O(kT(\Delta n+ \Delta m))$, where $\Delta n$ is number of candidate vertex in $cand$, and $\Delta m$ is total number of edges incident to vertices in $cand$. In practice, $\Delta n\ll n$ and $\Delta m\ll m \ll n^2$, therefore, we conclude that the computational cost per iteration is significantly reduced using Algorithm~\ref{alg::incremental-mu} than using the global approach. The time complexity of computing the transition matrix $A$ using Algorithm~\ref{alg::incremental-mu} is same with that of using the global approach.

\subsection{Real-time forecasting}\label{sec-batch}
In this section, we discuss how to apply our learning algorithms to real-time traffic prediction, where the sensor reading is received in a streaming fashion. In practice, if we want to make a prediction for the current traffic, we cannot afford to apply our global learning algorithm to all the previous snapshots because it is computationally expensive. Moreover, it is not always true that more snapshots would yield a better prediction performance. The alternative method is to treat each snapshot independently: i.e., each time we only apply our learning algorithm for the most recent snapshot, and then use the learned latent attribute to predict the traffic condition. Obviously, this would yield poor prediction quality as it totally ignores the temporal transitions.

 \begin{figure}[!h]
 	\centering
 	\includegraphics[width=75mm]{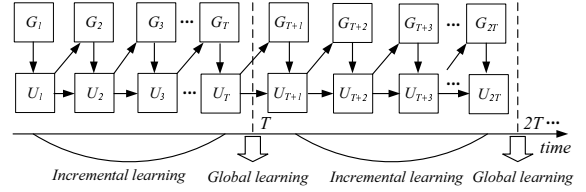}
 	\vspace{-3mm}\caption{A batch recomputing framework for real-time forecasting. To provide a just-in-time prediction, at each time span (a time span is very short, e.g., 5 minutes), we train our traffic model with the proposed incremental approach on-the-fly, and return the prediction instantly. Meanwhile, in order to maintain good quality, we batch recompute our traffic model per time window (a time window is much longer, e.g., one hour). } \vspace{-2mm} \label{fig::batch}
 \end{figure} 
 
To achieve a good trade-off between the above two methods, we propose to adapt a sliding window-based setting for the learning of our LSM-RN model, where we apply incremental algorithm at each timestamp during one time window, and only run our global learning algorithm at the end of one time window. As shown in Figure~\ref{fig::batch}, we apply our global learning at timestamps $T$ (i.e., the end of one time window), which learns the time-dependent latent attributes for the previous $T$ timestamps. Subsequently, for each timestamp $T+i$ between [T, 2T], we apply our incremental algorithm to adjust the latent attribute and make further predictions: i.e., we use $U_{T+i}$ to predict the traffic of $G_{T+i+1}$. Each time we receive the true observation of $G_{T+i+1}$, we calculate $U_{T+i+1}$ via the incremental update from Algorithm~\ref{alg::incremental-update}.  The latent attributes will be re-computed at timestamp $2T$ (the end of one time window), and the recomputed latent attributes $U_{2T}$ would be used for the next time window $[2T, 3T]$. 

\section{Experiment}\label{sec-exp}
\subsection{Dataset}
We used a large-scale and high resolution (both spatial and temporal) traffic sensor (loop detector) dataset collected from  Los Angeles County highways and arterial streets. This dataset includes both inventory and real-time data for 15000 traffic sensors covering approximately 3420 miles. The sampling rate of the streaming data, which provides speed, volume (number of cars passing from sensor locations) and occupancy,  is 1 reading/sensor/min. This sensor dataset have been continuously collected and archived since 2010.

\begin{figure}[!ht]
	\centering
	\includegraphics[width=.7\columnwidth]{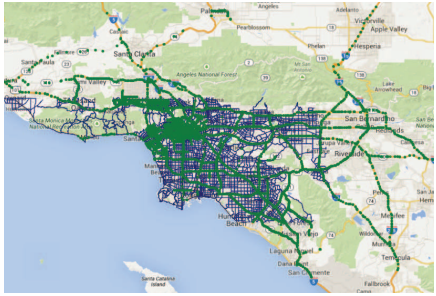}
	\vspace{-3mm}\caption{Sensor distribution and Los Angeles road network, where the green and blue lines depict the sensors and road network segments, respectively (Best viewed in color).} \vspace{-4mm} \label{fig::data-distribution}
\end{figure}

We chose two months of sensor dataset (i.e., March and April in 2014) for our experiments, which include more than 60 million records of readings. As for the road network, we used Los Angeles road network which was obtained from HERE Map dataset~\cite{navteq}. We constructed two subgraphs of Los Angeles road network, termed as SMALL and LARGE. The SMALL network contains 5984 vertices and 12538 edges, and LARGE contains 8242 vertices and 19986 edges. As described in Section~\ref{sec-problem}, the sensor data is mapped to the road network edges.
Figure~\ref{fig::data-distribution} shows sensors locations and road network segments, where the green lines depict the sensors, and blue lines represent the road network segments. 
After mapping the sensor data, we have two months of network snapshots for both SMALL and LARGE.

\subsection{Experimental setting}
\vspace{-1mm}
\subsubsection{Algorithms}
Our methods are termed as \textbf{LSM-RN-All} (i.e., global learning algorithm) and \textbf{LSM-RN-Inc} (i.e., incremental learning algorithm).



For edge traffic prediction,  we compare with LSM-RN-NAIVE, where we adapted the formulations from both~\cite{zhang2012overlapping} and~\cite{rossi2013modeling}. Hence, LSM-RN-NAIVE considers both network and temporal relationship, but uses the Naive incremental learning strategy as described in~\ref{}, which independently learns the latent attributes of each time-stamp first, then the transition matrix. We also compare our algorithms with two representative time series prediction methods: a linear model (i.e., ARIMA~\cite{PanICDM12}) and a non-linear model (i.e., SVR~\cite{RistPAKDD13}). We train each model independently for each time series with historical data. In addition, because these methods will be affected negatively due to the missing values during the prediction stages (i.e, some of the input readings for ARIMA and SVR could be zero), for fair comparison we consider ARIMA-Sp and SVR-Sp, which use the completed readings from our global learning algorithm. 

We consider the task of missing-value and missing sensor completion.
For the task of missing-value completion, we compare our algorithms with two methods: (1) KNN~\cite{haworth2012non}, which uses the average values of the nearby edges in Euclidean distance as the imputed value, (2) LSM-RN-NAIVE, which independently learns the latent attributes of each snapshot, then uses them to approximate the edge readings. We also implemented the Tensor method~\cite{TTB_Software,TTB_CPWOPT} for missing value completion. However, it cannot address the sparsity problem of our dataset and thus produce meaningless results (i.e., most of the completed values are close to 0). Although our framework is very general and supports missing sensor completion, we do not evaluate it through our experiments since we do not have ground truth values to verify.


To evaluate the performance of online prediction, we consider the scenario of a batch-window setting described in Section~\ref{sec-batch}. Considering a time window $[0, 2T]$, we sequentially predict the traffic condition for the timestamps during $[T+1, 2T]$, with the latent attributes of $U_T$ and transition matrix $A$ learned from the previous batch computing.
Each time when we make a prediction, we receive the true observations as the feedback. We compare our Incremental algorithm (Inc), with three baseline algorithms: Old, LSM-RN-Naive and LSM-RN-ALL. Specifically, in order to predict $G_{T+i}$, LSM-RN-Inc utilizes the feedback of $G_{T+i-1}$ to adjust the time-dependent latent attributes of $U_{T+i-1}$, whereas Old does not consider the feedback, and always uses latent attributes $U_T$ and transition matrix $A$ from the previous time window. 
On the other hand, LSM-RN-Naive ignores the previous snapshots, and only applies the inference algorithm to the most recent snapshot $G_{T+i-1}$ (aka Mini-batch). Finally, LSM-RN-All applies the global learning algorithm consistently to all historical snapshots (i.e., $G_1$ to $G_{T+i-1}$) and then makes a prediction (aka Full-batch).

\vspace{-2mm}
\begin{table}[!htbp]
	\centering
	\caption{\small {Experiment parameters}}\label{tbl:setting}
	\scalebox{0.80}{
		\begin{tabular}{|l|l|}
			\hline
			Parameters&Value range\\
			\hline
		$T$ & $2,4,6,8,\mathbf{10},12$\\
			$span$ & $\mathbf{5},10,15,20,25,30$ \\
			$k$ &  $5,10,15,\mathbf{20},25,30$\\
			$\lambda$  & $2^{-7},2^{-5},2^{-3},2^{-1}, 2^{1},\mathbf{2^{3}},2^{5}$\\
			$\gamma$  & $ 2^{-7}, \mathbf{2^{-5}},2^{-3},2^{-1},2^{1},2^{3},2^{5}$\\
			\hline	
		\end{tabular}}
		\vspace{-2mm}
	\end{table}

\subsubsection{Configurations and measures.}
With our missing value completion and edge traffic prediction experiments, we selected two different time ranges that represent rush hour (i.e., 7am-8am) and non-rush hour (i.e., 2pm-3pm) respectively. For the task of missing value completion, during each timestamps of one time range (e.g., rush hour), we randomly selected $20\%$ of values as unobserved and manipulated them as missing, with the objective of completing those missing values. For each traffic prediction task at one particular timestamp (e.g., 7:30 am), we also randomly selected $20\%$ of the values as unknown and use them as ground-truth values.


We varied the parameters $T$ and $span$: where $T$ is the number of snapshots, and $span$ is time gap between two continuous snapshots. We also varied $k$, $\lambda$, and $\gamma$, which are parameters of our model. The default settings (shown with bold font) of the experiment parameter are listed in Table~\ref{tbl:setting}. Because of space limitations, the results of varying $\gamma$ are not reported, which are similar to result of varying $\lambda$.
We use Mean Absolute Percentage Error (MAPE) and Root Mean Square Error (RMSE) to measure the accuracy. In the following we only report the experiment results based on MAPE, the experiment results based on RMSE are reported in Appendix~\ref{sec::app-3}. Specifically, MAPE is defined as follows:


\begin{small}
\begin{equation*}
\begin{aligned}
MAPE = (\frac{1}{N}\sum_{i=1}^{N}\frac{|y_i-\hat{y_i}|}{y_i})
\end{aligned}
\end{equation*}
\end{small}

With ARIMA and SVR, we use the dataset of March to train a model for each edge, and use 5-fold cross-validation to choose the best parameters. All the tasks of missing value completion and edge traffic prediction tasks are conducted on April data. We conducted our experiments with C++ on a Linux PC with  i5-2400 CPU @ 3.10G HZ and 24GB memory.

\begin{figure}[!ht]
	\begin{small}\scalebox{0.99}{
			\begin{tabular}{cc}
				\multicolumn{2}{c}{\hspace{-9mm} \includegraphics[width = 82mm]{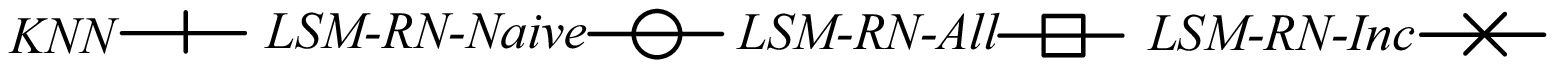}}
				\vspace{-1mm} \\
				\hspace{-15mm}\includegraphics[width=55mm]{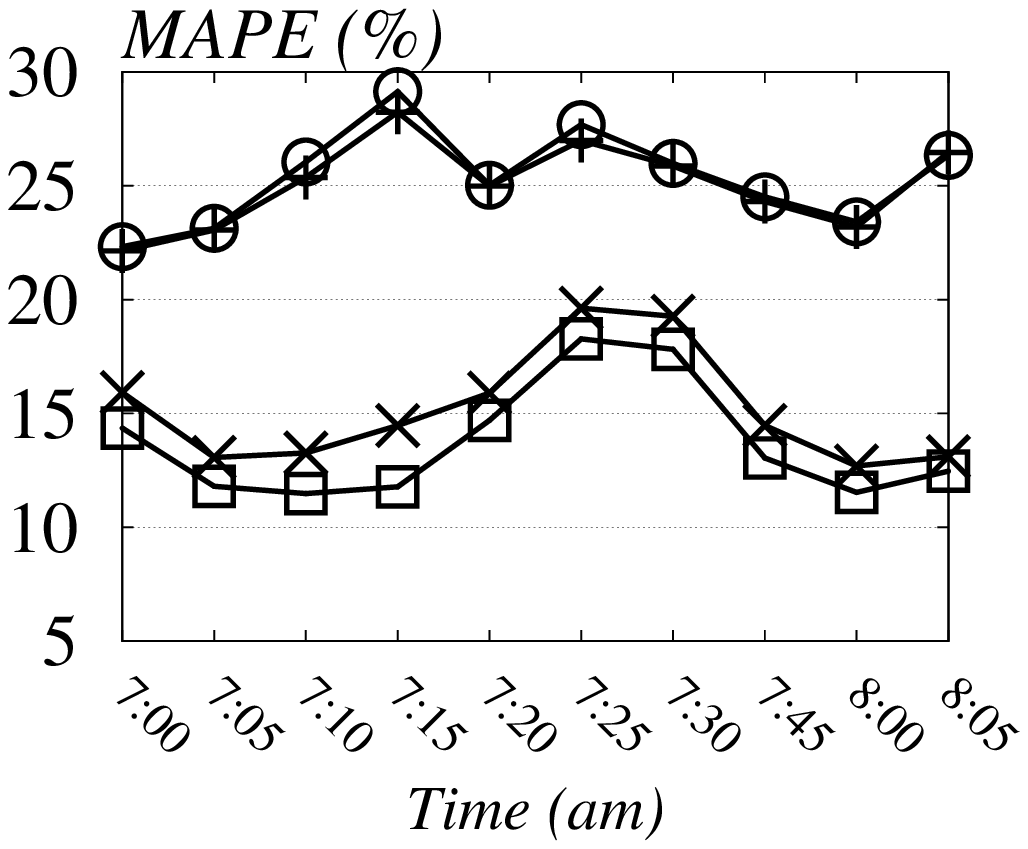} &
				\vspace{-2mm}\hspace{-15mm}\includegraphics[width=55mm]{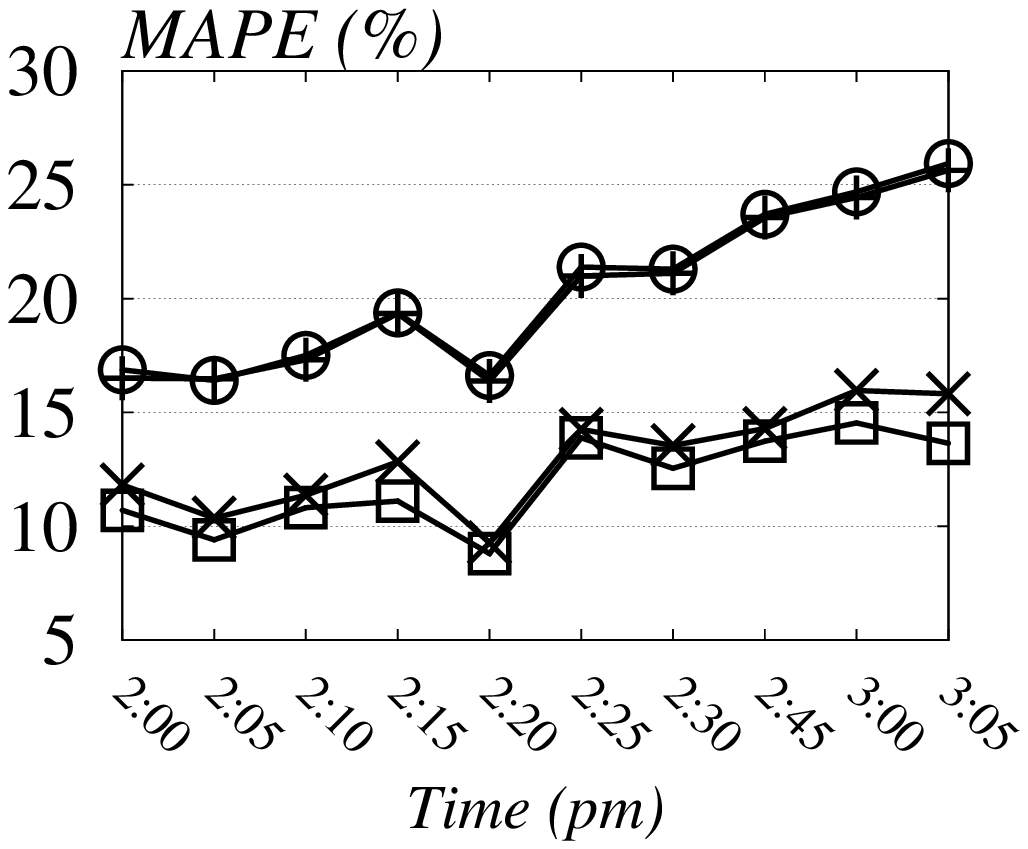}  \vspace{1mm}\\
				\hspace{-6mm}(a) Rush hour on SMALL & \hspace{-5mm}(b) Non-Rush hour on SMALL\\
				\hspace{-15mm}\includegraphics[width=55mm]{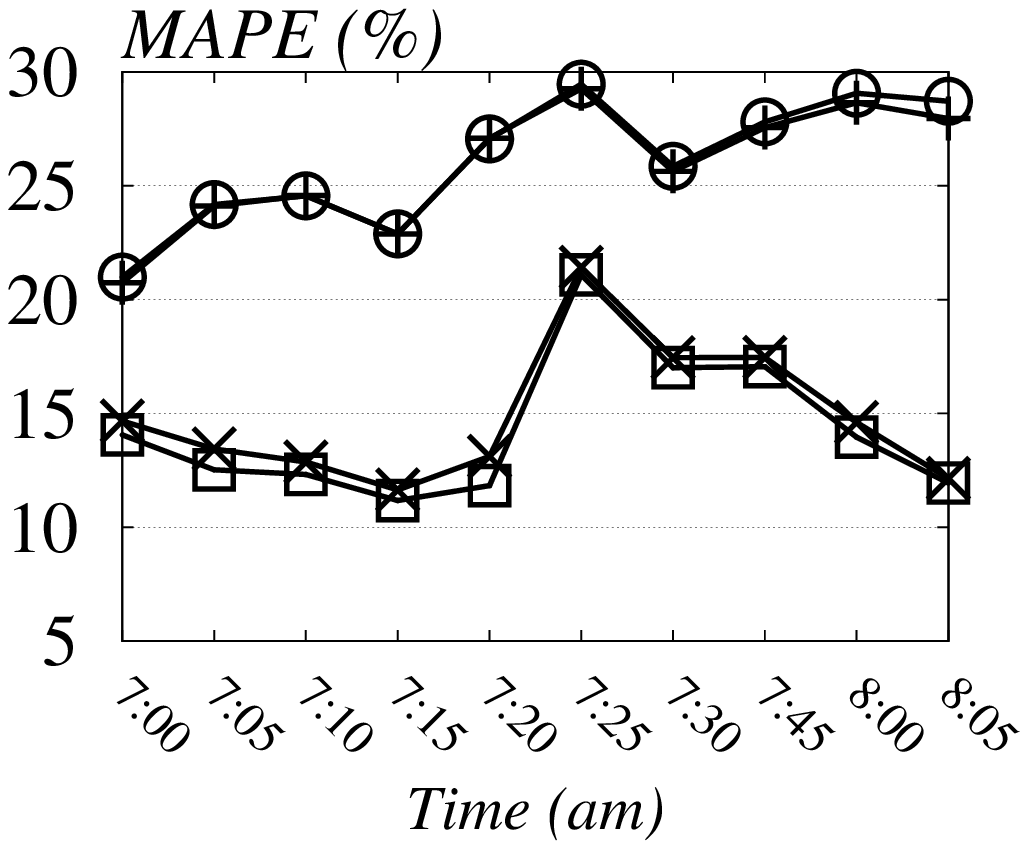} &
				\vspace{-2mm}\hspace{-15mm}\includegraphics[width=55mm]{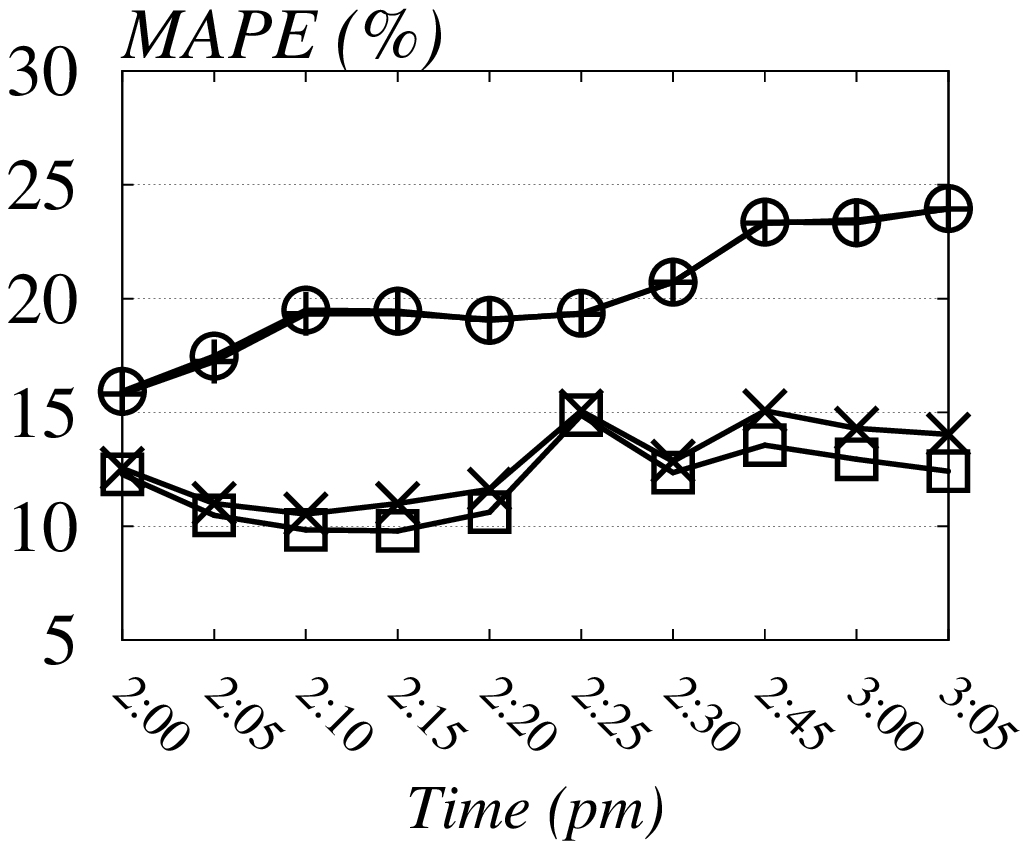}  \vspace{1mm}\\	
				\hspace{-6mm}(c) Rush hour on LARGE & \hspace{-5mm}(d) Non-Rush hour on LARGE
			\end{tabular}}
		\end{small}
		\vspace{-3mm}\caption{\small Missing value completion MAPE } \vspace{-0mm}		
		\vspace{-0mm}\label{fig::exp-comp-mape}
\end{figure}\vspace{-0mm}
	
\subsection{Comparison for missing value completion}
In this set of experiments, we evaluate the completion accuracy of different methods. The experiment results on SMALL are shown in Figure~\ref{fig::exp-comp-mape} (a) and (b). We observe that both LSM-RN-All and LSM-RN-Inc achieve much lower errors than that of other methods. This is because LSM-RN-All and LSM-RN-Inc capture both spatial and temporal relationships, while LSM-RN-One and KNN only use spatial property. LSM-RN-All performs better than LSM-RN-Inc by jointly infer all the latent attributes. On the other hand, we note that LSM-RN-One and KNN have similar performances, which shows that the effect of utilizing either Euclidean or Topology proximity is not enough for missing value completion. 
 This also indicates that utilizing both spatial and temporal property yields a large gain than only utilizing spatial property.

As shown in Figure~\ref{fig::exp-comp-mape}(b), the completion performance on the non-rush hour is better as compared to on the rush hour time interval. This is because during  rush hour range, the traffic condition is more dynamic, and the underlying pattern and transition changes frequently. All of these factors render worse performance during rush hour. Figure~\ref{fig::exp-comp-mape} (c) and (d) depict the experiment results on LARGE, which are similar on that of SMALL.

\vspace{-1mm}
\subsection{Comparison with edge traffic prediction}
In the following, we present the results of edge traffic prediction experiments.

\begin{figure}[!ht]
	\begin{small}\scalebox{0.99}{
			\begin{tabular}{cc}
				\multicolumn{2}{c}{	\vspace{+1mm}\hspace{-12mm} \includegraphics[width = 70mm]{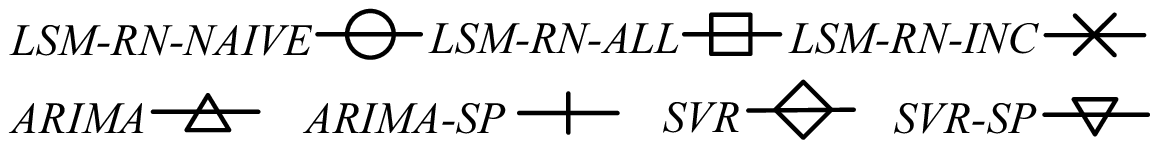}}
				\vspace{-1mm} \\
				\hspace{-9mm}\includegraphics[width=50mm]{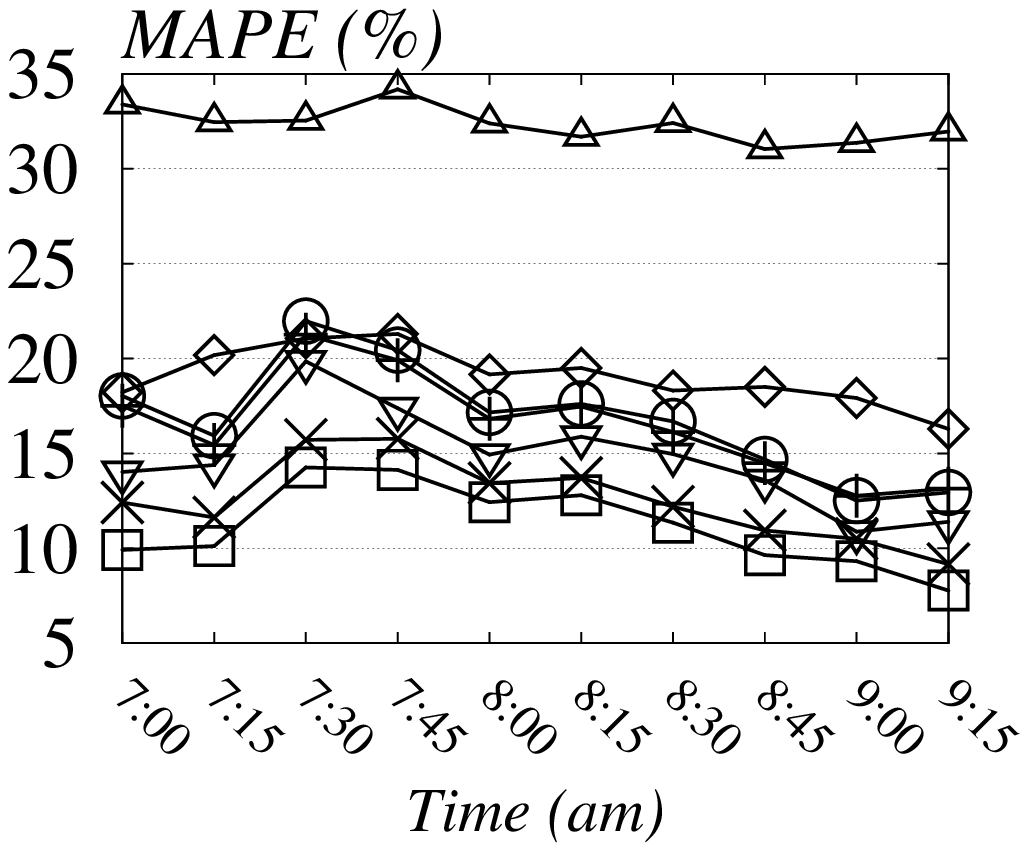} &
				\vspace{-2mm}\hspace{-9mm}\includegraphics[width=50mm]{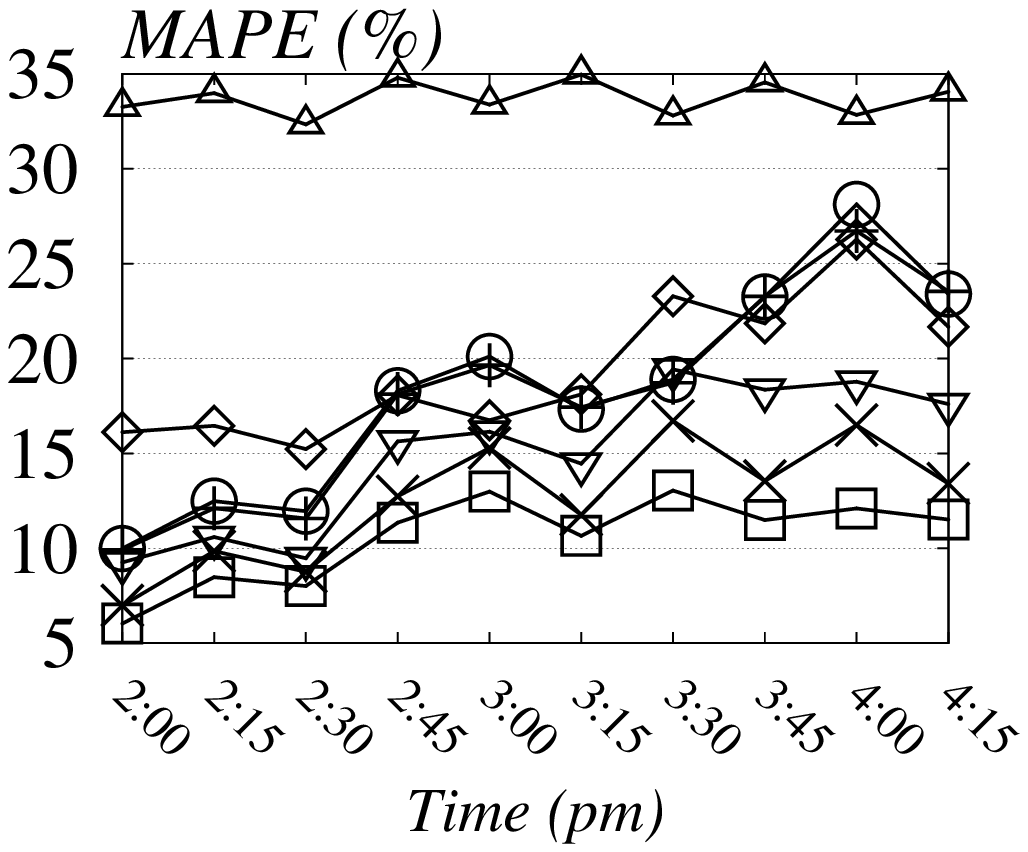}  \vspace{1mm}\\
				\hspace{-6mm}(a) Rush hour on SMALL & \hspace{-5mm}(b) Non-Rush hour on SMALL\\
				\hspace{-9mm}\includegraphics[width=50mm]{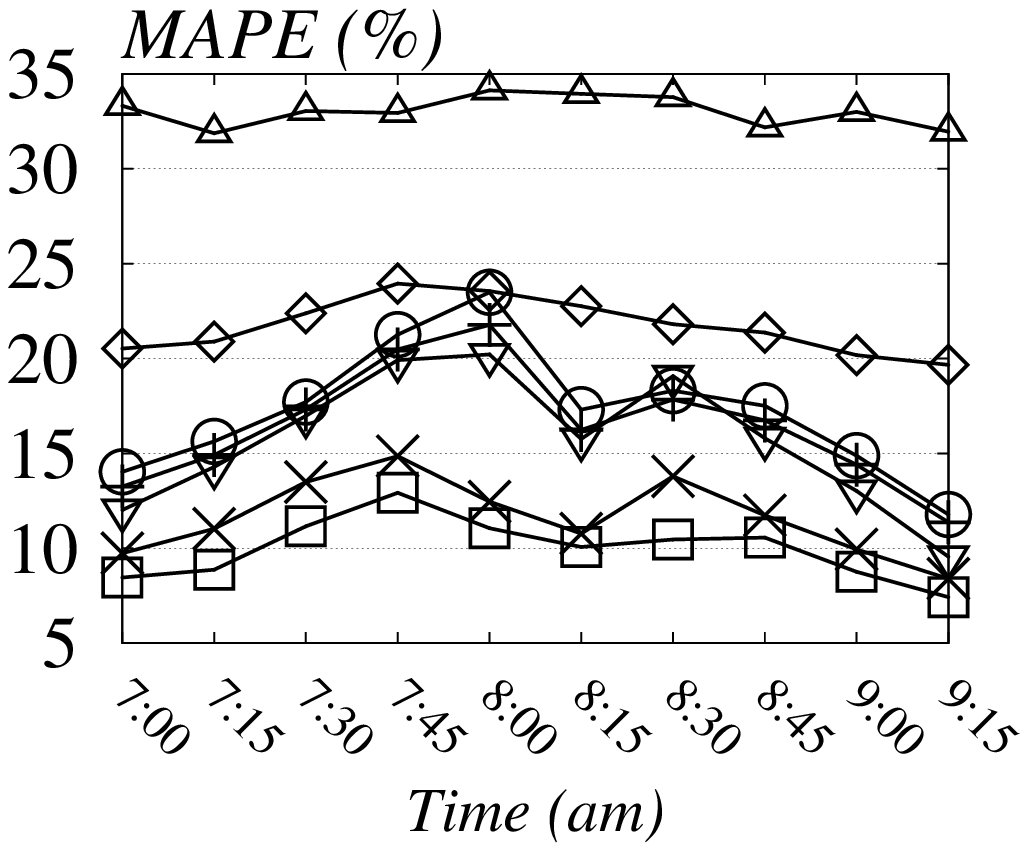} &
				\vspace{-2mm}\hspace{-9mm}\includegraphics[width=50mm]{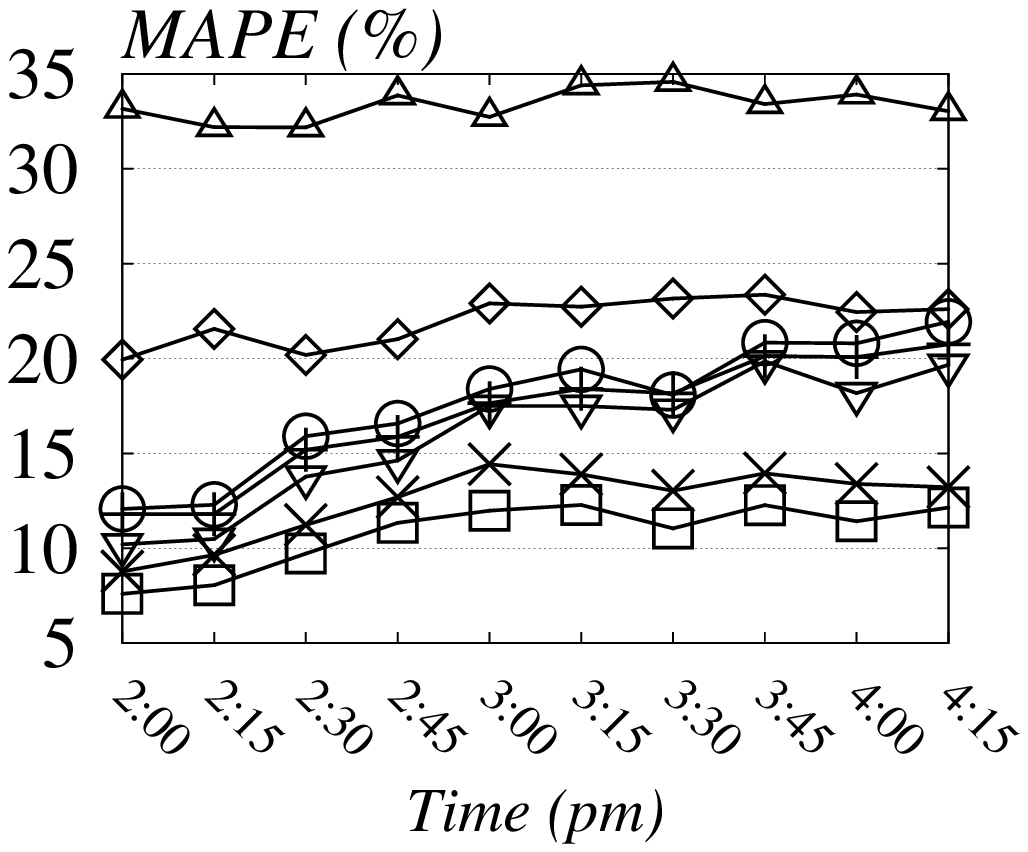}  \vspace{1mm}\\
				\hspace{-6mm}(c) Rush hour on LARGE & \hspace{-5mm}(d) Non-Rush hour on LARGE
			\end{tabular}}
		\end{small}
		\vspace{-3mm}\caption{\small One-step ahead prediction MAPE} \vspace{-0mm}
		\label{fig::exp-pred-mape}
		\vspace{-0mm}
\end{figure}\vspace{-0mm}	
			
\subsubsection{One-step ahead prediction}
The experimental results of SMALL are shown in Figure~\ref{fig::exp-pred-mape} (a) and (b). Among all the methods, LSM-RN-All and LSM-RN-Inc achieve the best results, and LSM-RN-All performs slightly better than LSM-RN-Inc. This demonstrates that the effectiveness of time-dependent latent attributes and the transition matrix. 
We observe that without the imputation of missing values, time series prediction techniques (i.e., ARIMA and SVR) perform much worse than LSM-RN-ALL and LSM-RN-Inc. Meanwhile, LSM-RN-Naive, which separately learns the latent attributes of each snapshot, cannot achieve good prediction results as compared to LSM-RN-All and LSM-RN-Inc. This indicates that simply combining topology and temporal is not enough for making accurate predictions. We also note that even with completed readings, the accuracy of SVR-Sp and ARIMA-Sp is still worse than that of LSM-RN-All and LSM-RN-Inc. One reason is that simply combination of the spatial and temporal properties does not necessarily yield a better performance. Another reason is that both SVR-Sp and ARIMA-Sp also suffer from missing data during the training stage, which renders less accurate prediction. In Appendix~\ref{sec::app-4}, we also show how the ratio of missing data would influence the prediction performance. Finally, we observe that SVR is more robust than ARIMA when encountering missing value on prediction stages: i.e., ARIMA-Sp performs significantly better than ARIMA, while the improvement of SVR-Sp over SVR is not much. This is because ARIMA is a linear model which mainly uses the weighted average of the previous readings for prediction, while SVR is a non-linear model that utilizes a kernel function. Figure~\ref{fig::exp-pred-mape} (c) and (d) show the experiment results on LARGE, the overall results are similar to those of SMALL.




\begin{figure}[!ht]
	\begin{small}\scalebox{0.99}{
			\begin{tabular}{cc}
				\multicolumn{2}{c}{\vspace{+1mm}\hspace{-12mm} \includegraphics[width = 70mm]{fig/Exp-Legends-new}}
				\vspace{-1mm} \\
				\hspace{-9mm}\includegraphics[width=50mm]{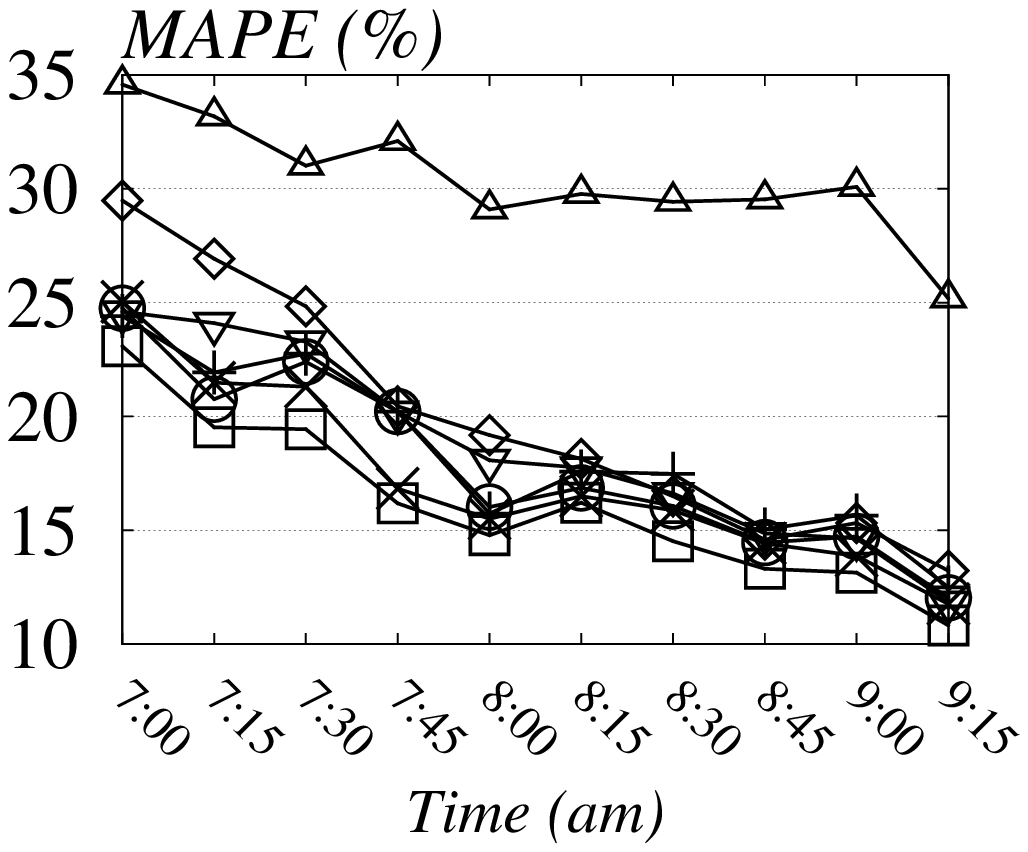} &
				\vspace{-2mm}\hspace{-9mm}\includegraphics[width=50mm]{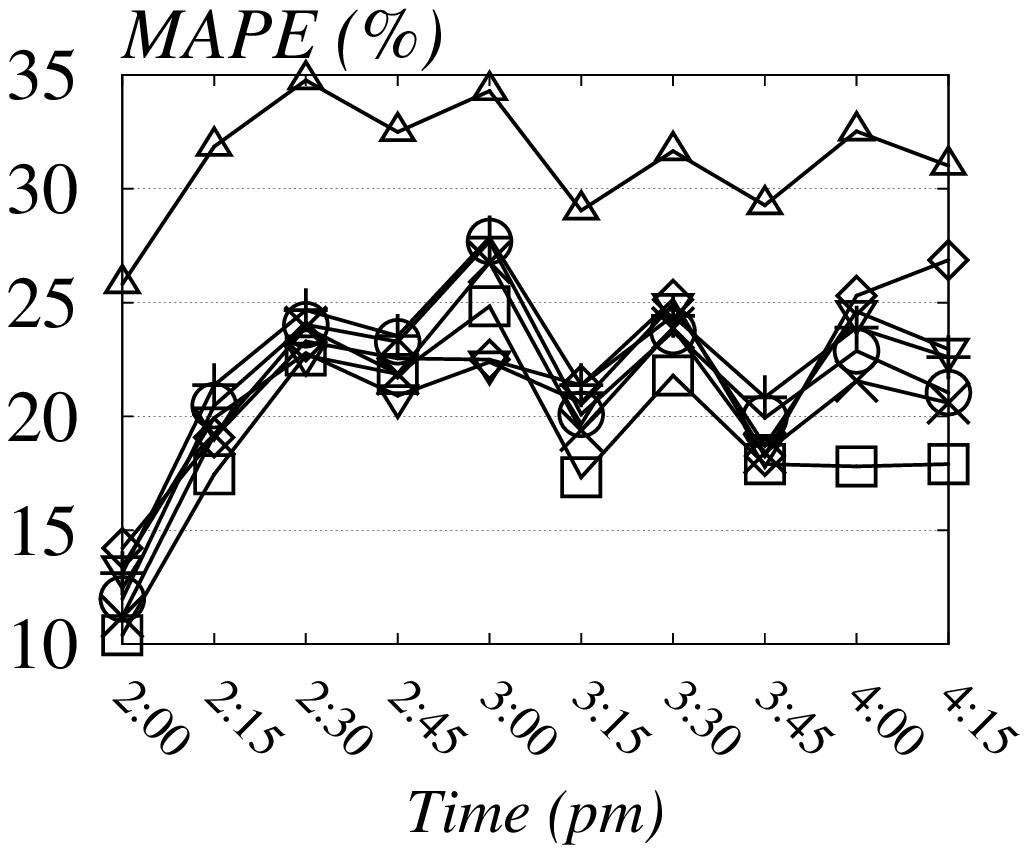}  \vspace{1mm}\\
				\hspace{-6mm}(a) Rush hour on SMALL & \hspace{-5mm}(b) Non-Rush hour on SMALL\\
				\hspace{-9mm}\includegraphics[width=50mm]{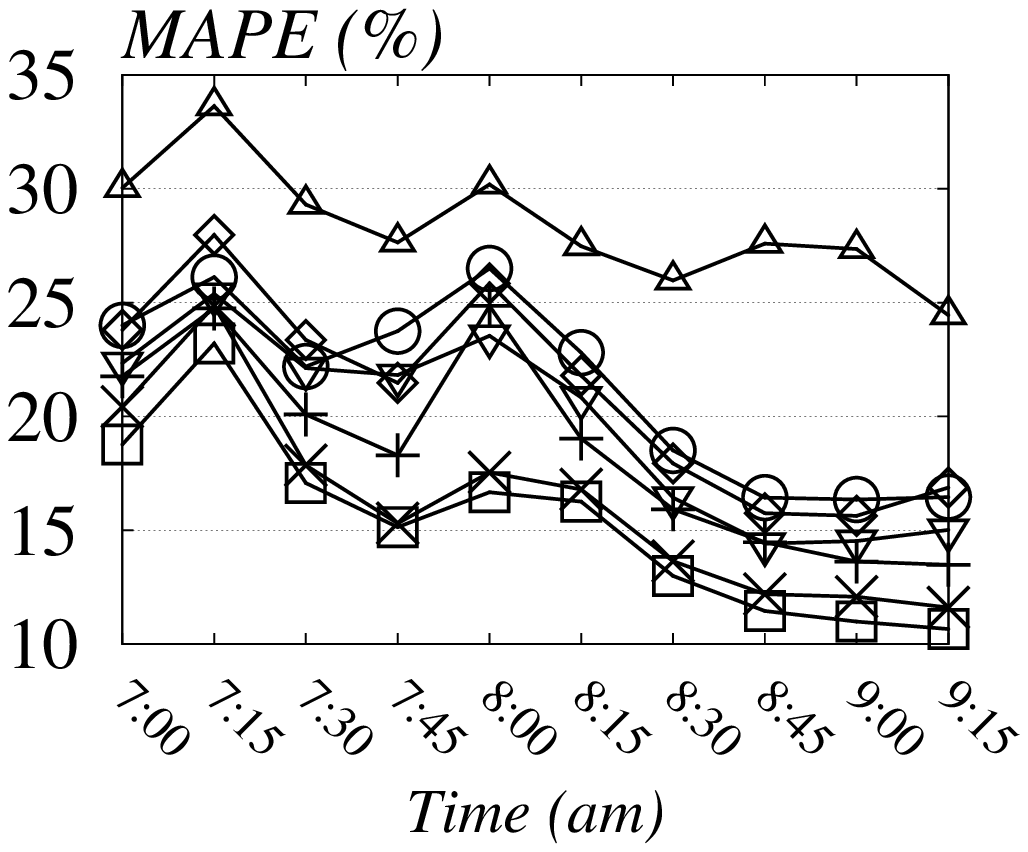} &
				\vspace{-2mm}\hspace{-9mm}\includegraphics[width=50mm]{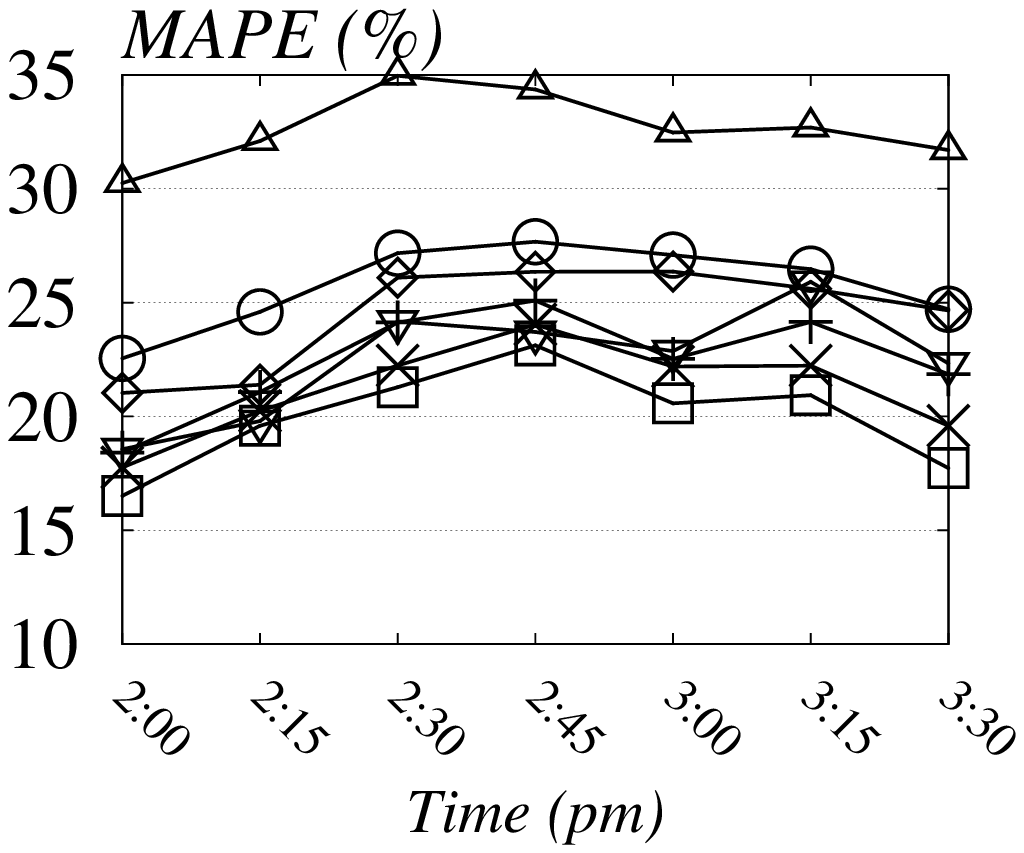}  \vspace{1mm}\\
				\hspace{-6mm}(c) Rush hour on LARGE & \hspace{-5mm}(d) Non-Rush hour on LARGE
			\end{tabular}}
		\end{small}
		\vspace{-3mm}\caption{\small Six-steps ahead prediction MAPE} \vspace{-0mm}
		\label{fig::exp-pred-mult-mape}
		\vspace{-0mm}
\end{figure}\vspace{-0mm}

\subsubsection{Multi-steps ahead prediction}
We now present the experiment results on multi-step ahead prediction, with which we predict the traffic conditions for next 30 minutes (i.e., $h=6$).
The prediction accuracy comparison among different methods on SMALL are shown in Figure~\ref{fig::exp-pred-mult-mape} (a) and (b). Although LSM-RN-All and LSM-RN-Inc still outperforms other methods, the margin between our methods and the baselines is smaller. The reason is that, when we make multiple-step ahead prediction, we use the predicted values from the past for future prediction. This leads to the problem of error accumulation, i.e., errors incurred in the past are propagated into future predictions. We observe the similar trends on LARGE, from the results reported in Figure~\ref{fig::exp-pred-mult-mape} (c) and (d).



\subsection{Scalability of different methods}
Table~\ref{tab::runtime} shows the running time of different methods. Although ARIMA and SVR is fast for each prediction, they have much higher training cost. Note that our methods do not require extra training data, i.e., our methods train and predict at the same time. Among them, LSM-RN-Inc is the most efficient approach: it only takes less than 500 milliseconds to learn the time-dependent latent attributes and make predictions for all the edges of the road network.
This is because our incremental learning algorithm conditionally adjusts the latent attributes of certain vertices, and utilizes the topological order that enables fast convergence. Even for the LARGE dataset, LSM-RN-Inc only takes less than five seconds, which is acceptable considering the span between two snapshots is at least five minutes in practice. This demonstrates that LSM-RN-Inc scales well to large road networks. Regarding LSM-RN-All and LSM-RN-Naive, they both require much longer running time than that of LSM-RN-Inc. In addition, LSM-RN-All is faster than LSM-RN-Naive. This is because LSM-RN-One independently runs the global learning algorithm for each snapshot $T$ times, while LSM-RN-All only applies global learning for the whole snapshots once.



\begin{table}[!ht]
 	\centering
 	\small{
 		\caption{Running time comparisons. For ARIMA and SVR, the training time cost is the total running time for training the models for all the edges for one-step ahead prediction in the experiment, and the prediction time  is the average prediction time per edge per query. For methods LSM-RN-Naive, LSM-RN-All and LSM-RN-Inc), we train (learn the latent attributes) and predict (the traffic condition of the whole road network) at the same time with the given data on-the-fly.}\label{tab::runtime}
 		\begin{tabular}{|c|c|c|c|c|}
 			\hline
 			data&\multicolumn{2}{c|}{SMALL}&\multicolumn{2}{c|}{LARGE}\\
 			\hline
 			&train (s)&pred.(ms)&train (s)&pred. (ms)\\
 			\hline
 			LSM-RN-One&-&1353&-&29439\\
 			\hline
 			LSM-RN-All&-&869&-&14247\\
 			\hline
 			LSM-RN-Inc&-&407&-&4145\\
 			\hline
 			\hline
 			ARIMA&484&0.00015&987&0.00024\\
 			\hline
 			SVR&47420&0.00042&86093.99&0.00051\\
 			\hline
 		\end{tabular}
 	}
\end{table}

\begin{figure}[!ht]
 	\begin{small}\scalebox{0.99}{
 			\begin{tabular}{cc}
 				\vspace{-1mm} \\
 				\hspace{-12mm}\includegraphics[width=50mm]{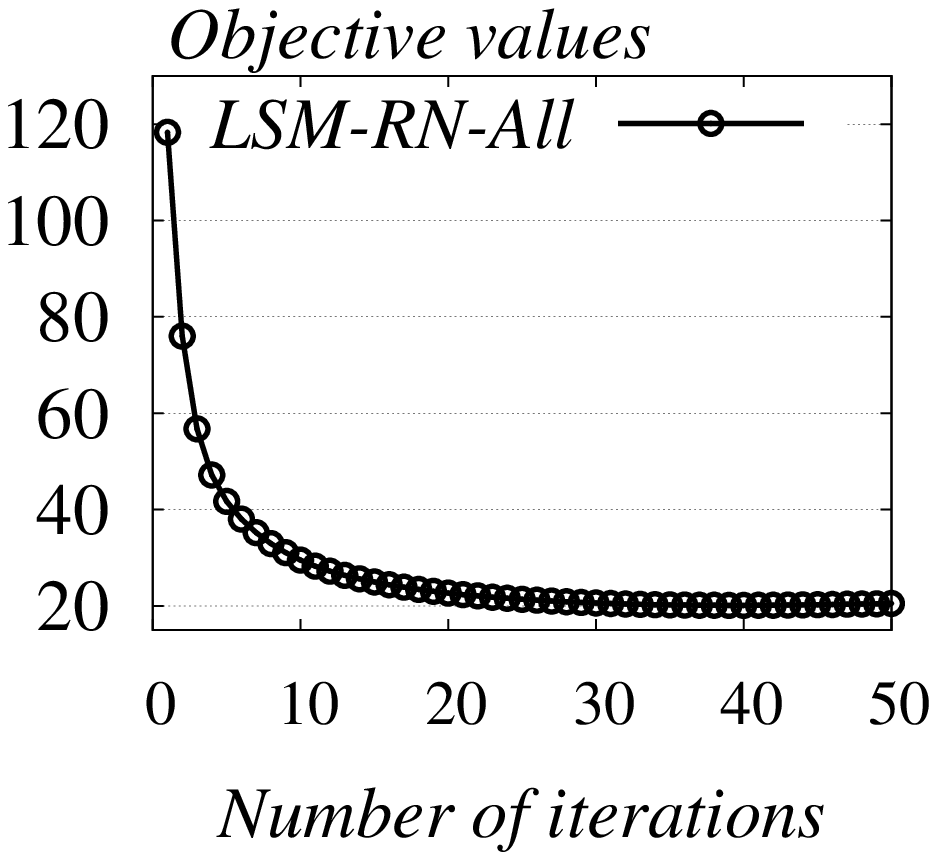} &
 				\vspace{-2mm}\hspace{-12mm}\includegraphics[width=50mm]{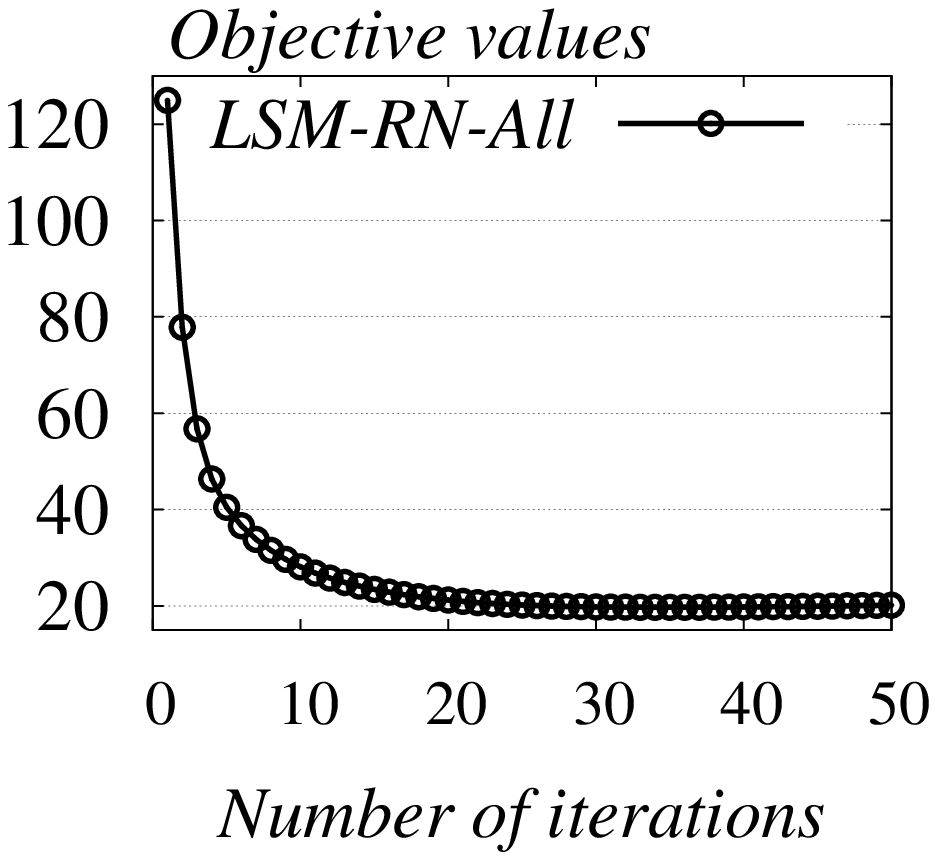}  \vspace{1mm}\\
 				\hspace{-6mm}(a) SMALL & \hspace{-5mm}(b) LARGE
 			\end{tabular}}
 		\end{small}
 		\vspace{-3mm}\caption{\small Converge rate} \vspace{-0mm}
 		\label{fig::exp-converge}
 		\vspace{-0mm}
 	\end{figure}

 \noindent\textbf{Convergence analysis.} Figure~\ref{fig::exp-converge} (a) and (b) report the convergence rate of iterative algorithm LSM-RN-All on both SMALL and LARGE. As shown in Figure~\ref{fig::exp-converge}, LSM-RN-All converges very fast: when the number of iterations is around 20, our algorithm tends to converge in terms of our objective value in Equation~\ref{eqn::final-obj}.


 \vspace{-0mm}

\subsection{Comparison for Real-time Forecasting}
\begin{figure}[!ht]
	\begin{small}\scalebox{0.99}{
			\begin{tabular}{cc}
				\multicolumn{2}{c}{\hspace{-5mm} \includegraphics[width = 82mm]{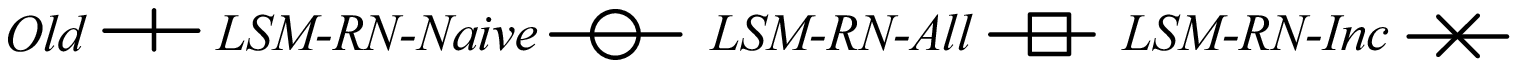}}
				\vspace{-1mm} \\
				\hspace{-12mm}\includegraphics[width=50mm]{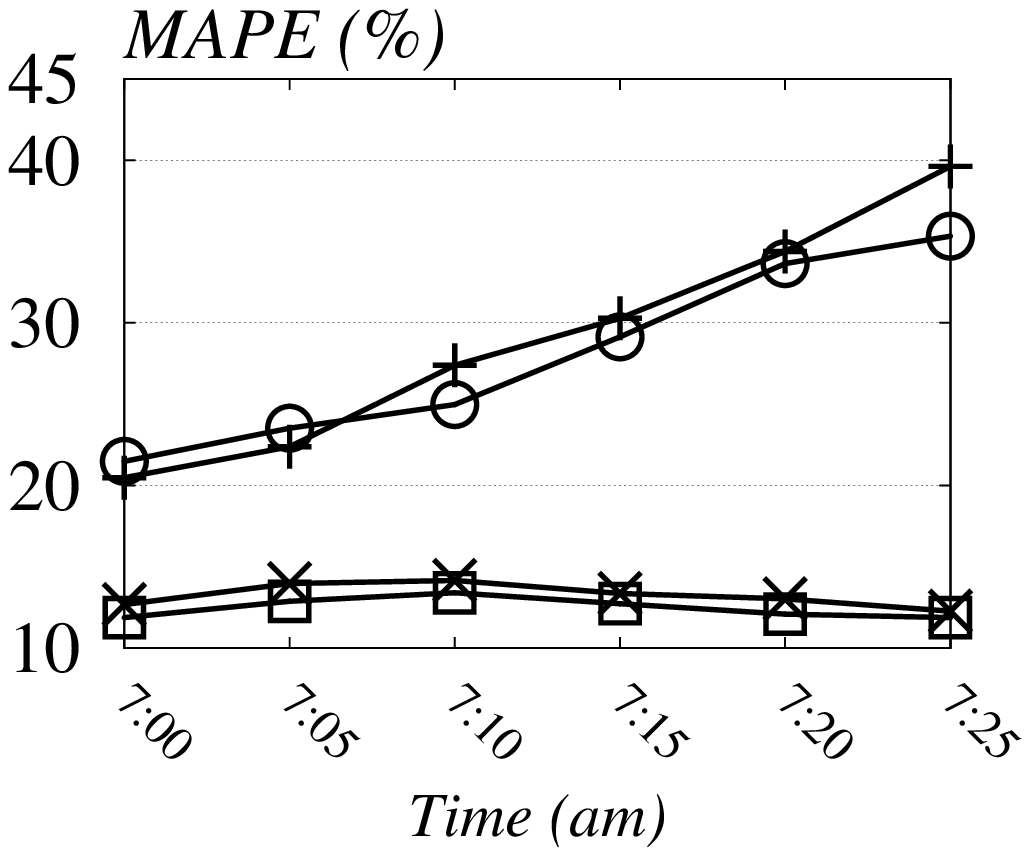} &
				\vspace{-2mm}\hspace{-12mm}\includegraphics[width=50mm]{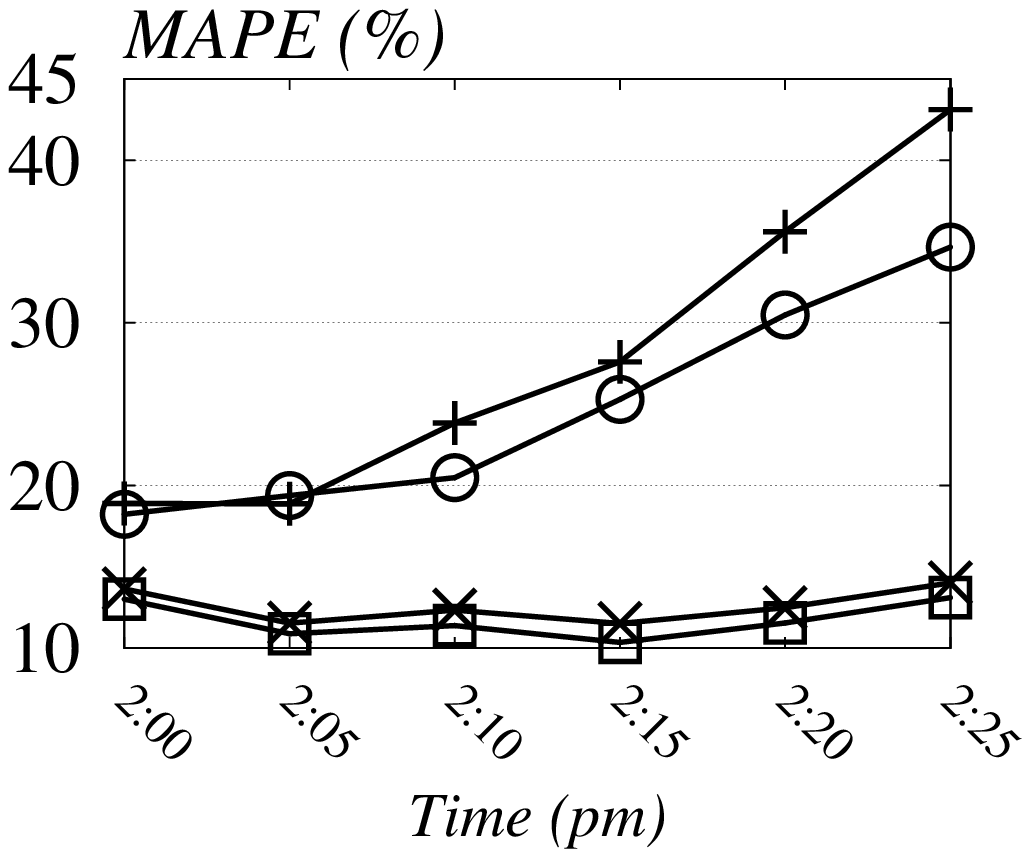}  \vspace{1mm}\\
				\hspace{-6mm}(a) Rush hour on SMALL & \hspace{-5mm}(b) Non-Rush hour on SMALL\\
				\hspace{-12mm}\includegraphics[width=50mm]{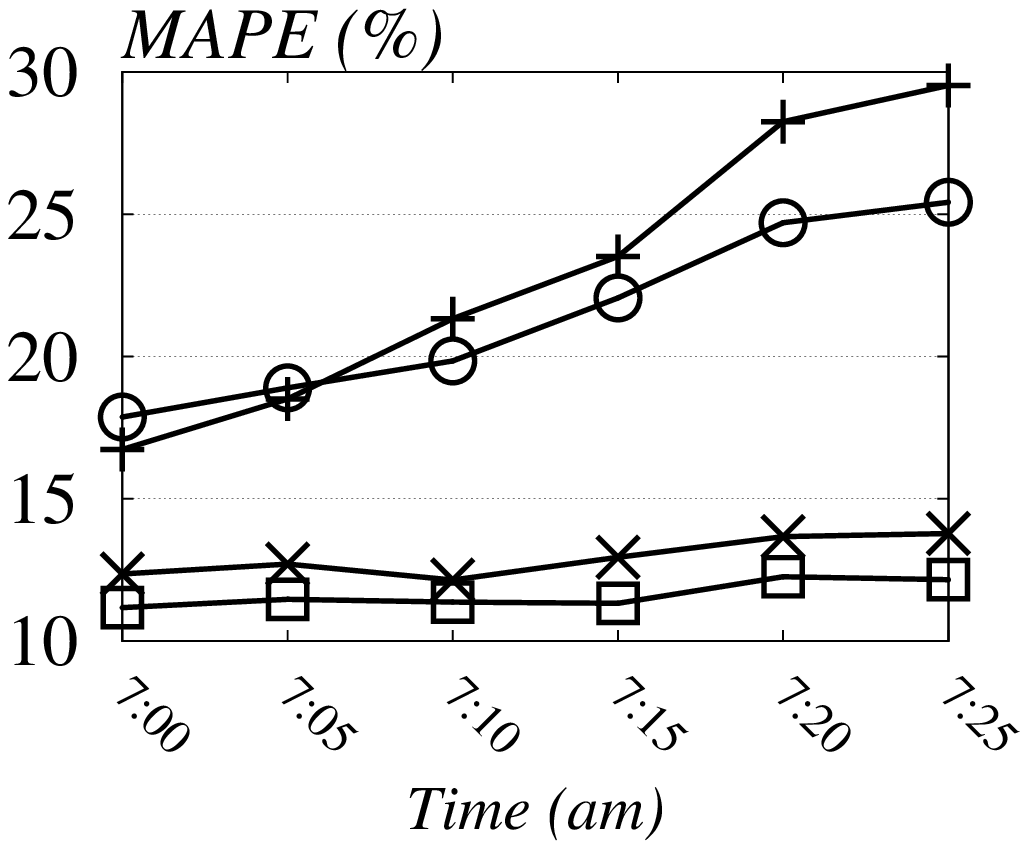} &
				\vspace{-2mm}\hspace{-12mm}\includegraphics[width=50mm]{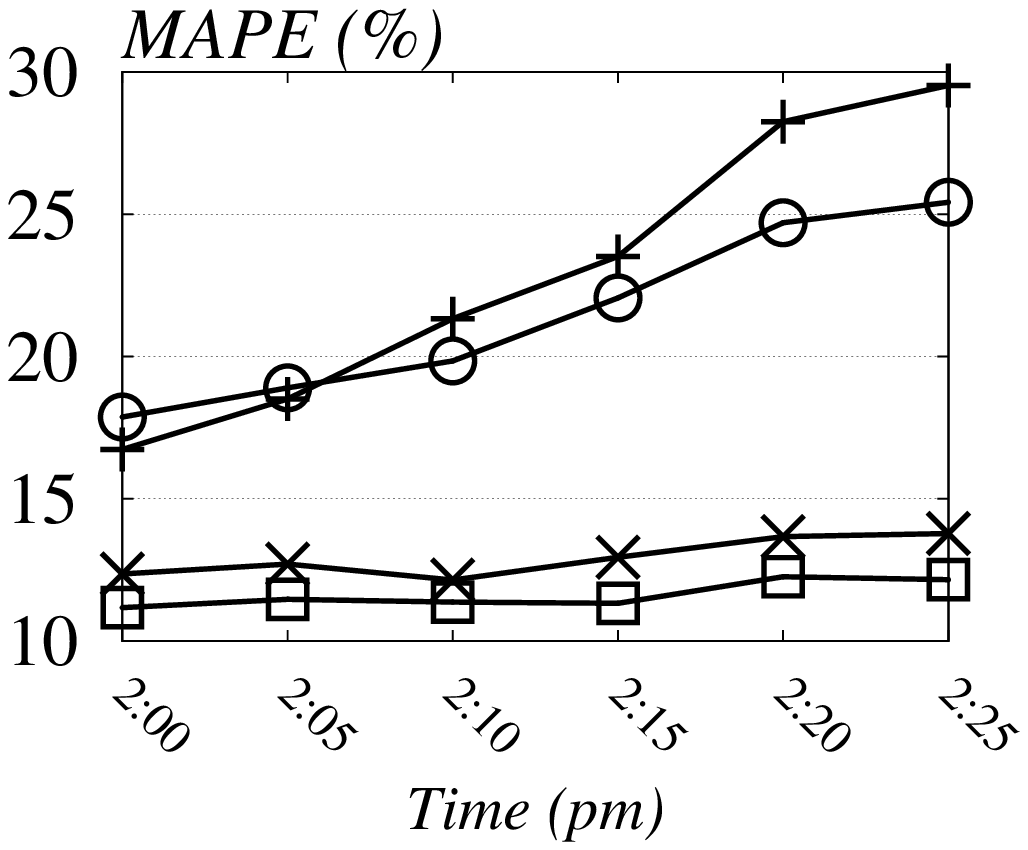}  \vspace{1mm}\\
				\hspace{-6mm}(c) Rush hour on LARGE& \hspace{-5mm}(d) Non-Rush hour on LARGE
			\end{tabular}}
		\end{small}
		\vspace{-3mm}\caption{\small Online prediction MAPE} \vspace{-0mm}
		\label{fig::exp-online-mape}
		\vspace{-0mm}
	\end{figure}

In this set of experiments, we evaluate our online setting algorithms. As shown in Figure~\ref{fig::exp-online-mape} (a) and (b), LSM-RN-Inc
achieves comparable accuracy with LSM-RN-ALL (Full-batch). This is because LSM-RN-Inc effectively leverages the real-time feedback information to adjust
the latent attributes. We observe that LSM-RN-Inc performs significantly better than Old and LSM-RN-Naive (Mini-batch), which ignores either the feedback information (i.e., Old) or the previous snapshots (i.e., LSM-RN-Naive). One interesting observation is that Old performs better than LSM-RN-Naive for the initial timestamps, whereas Old surpasses Mini-batch at the later timestamps. This indicates that the latent attributes learned in the previous time-window are more reliable for predicting the near-future traffic conditions, but may not be good for multi-step ahead prediction because of the error accumulation problem. Similar results have also been observed in Figure~\ref{fig::exp-pred-mult-mape} for multi-step ahead prediction. Figure~\ref{fig::exp-online-mape} (c) and (d) show similar effects on LARGE.

\begin{figure}[!ht]
	\begin{small}\scalebox{0.99}{
			\begin{tabular}{cc}
				\multicolumn{2}{c}{\hspace{-5mm} \includegraphics[width = 75mm]{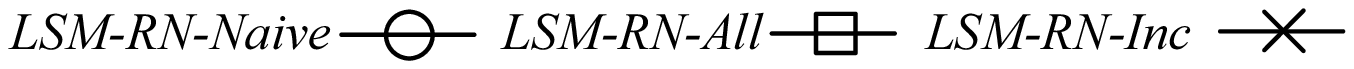}}
				\vspace{-1mm} \\
				\hspace{-12mm}\includegraphics[width=50mm]{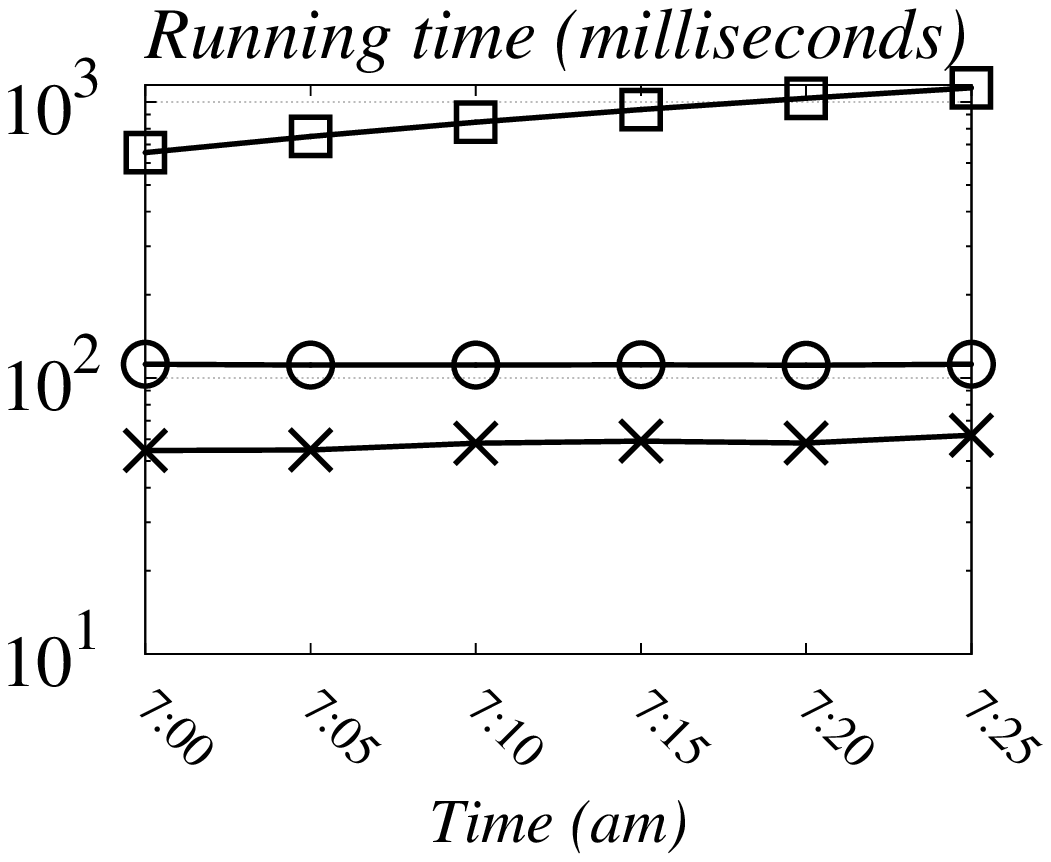} &
				\vspace{-2mm}\hspace{-12mm}\includegraphics[width=50mm]{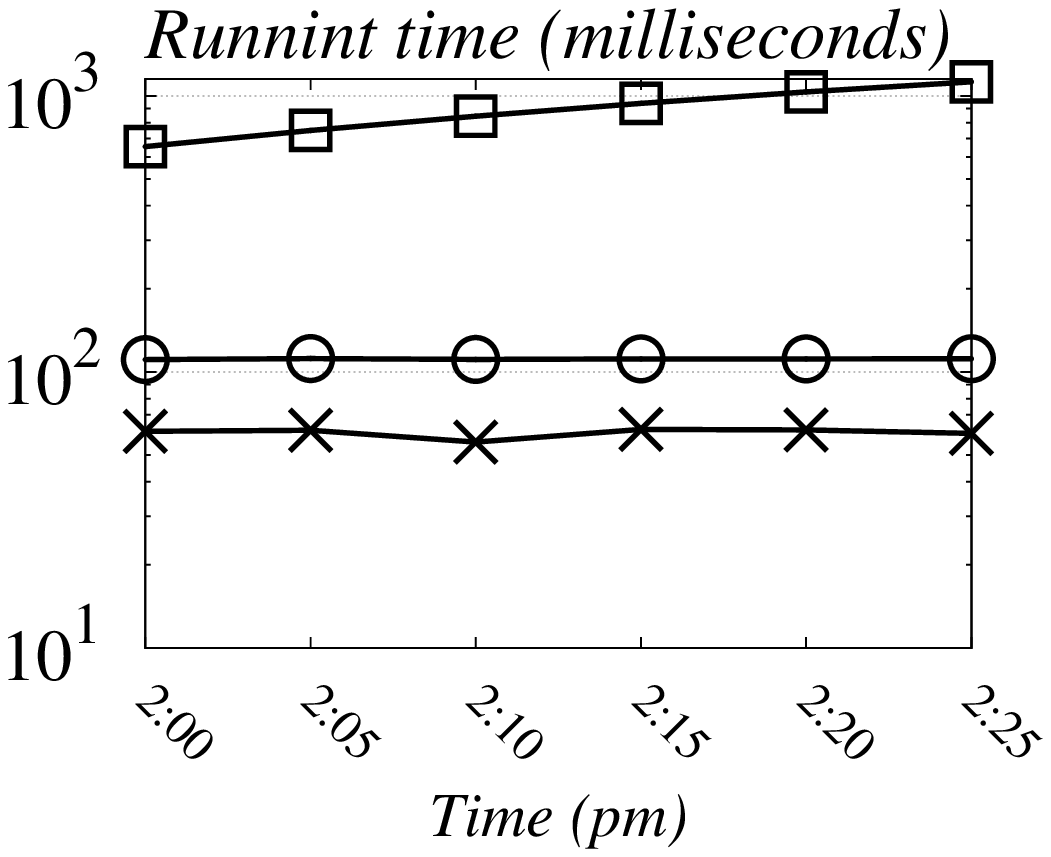}  \vspace{1mm}\\
				\hspace{-6mm}(a) Rush hour  on SMALL & \hspace{-5mm}(b) Non-Rush hour  on SMALL\\
					\hspace{-12mm}\includegraphics[width=50mm]{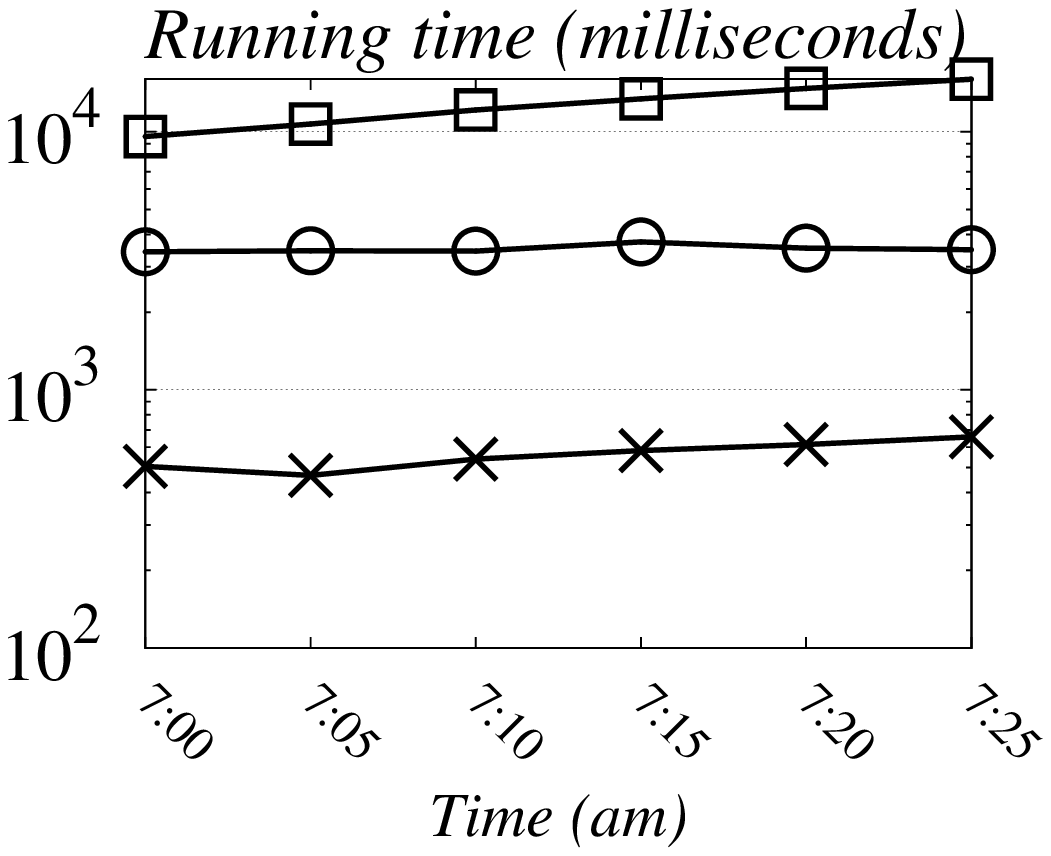} &
					\vspace{-2mm}\hspace{-12mm}\includegraphics[width=50mm]{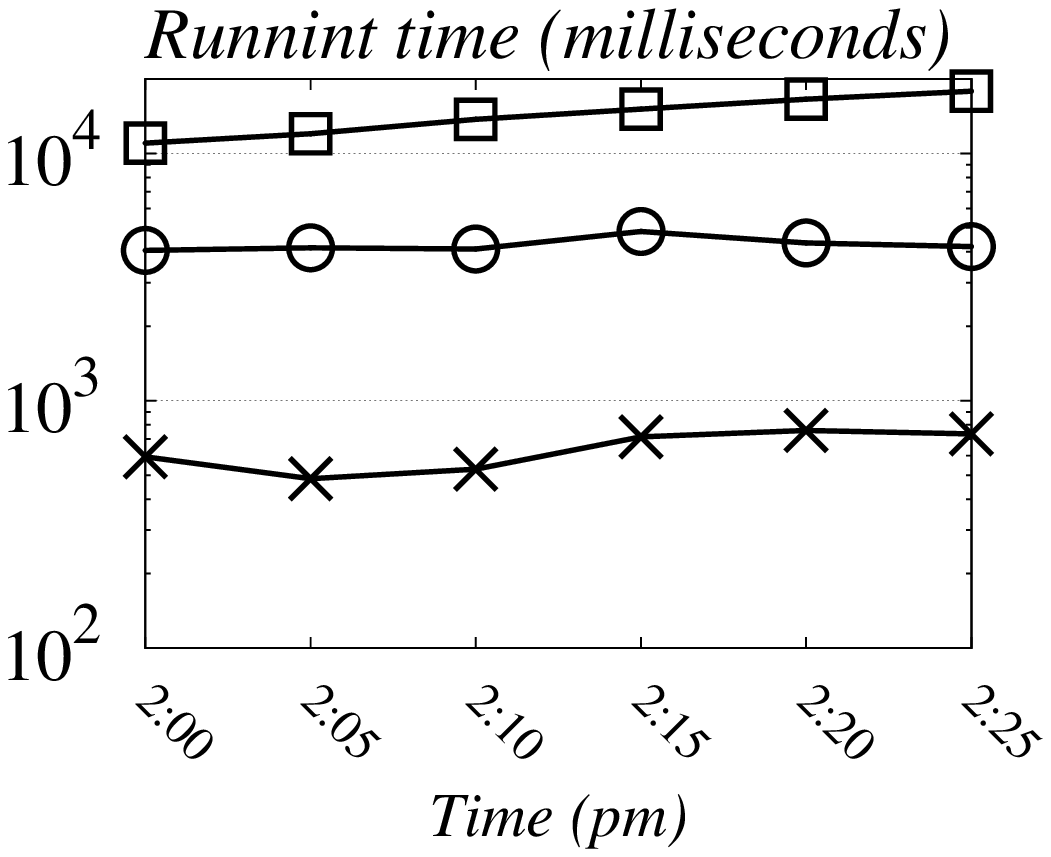}  \vspace{1mm}\\
					\hspace{-6mm}(c) Rush hour on LARGE & \hspace{-5mm}(d) Non-Rush hour  on LARGE
			\end{tabular}}
		\end{small}
		\vspace{-3mm}\caption{\small Online Prediction time} \vspace{-0mm}
		\label{fig::exp-online-time}
		\vspace{-0mm}
\end{figure}

Figure~\ref{fig::exp-online-time} (a) and (b) show the running time comparisons of different methods. One important conclusion for this experiment is that  LSM-RN-Inc is the most efficient approach, which is on average two times faster than LSM-RN-Naive and one order of magnitude faster than LSM-RN-ALL. This is because LSM-RN-Inc performs a conditional latent attribute update for vertices within a small portion of road network, whereas LSM-RN-Naive and LSM-RN-All both recompute the latent attributes from at least one entire road network snapshot. Because in the real-time setting, LSM-RN-All utilizes all the up-to-date snapshots and LSM-RN-Naive only considers the most recent single snapshot, LSM-RN-Naive is faster than LSM-RN-All. Figure~\ref{fig::exp-online-time} (c) and (d) show the running time on LARGE. We observe that LSM-RN-Inc only takes less than 1 seconds to incorporate the real-time feedback information, while LSM-RN-Naive and LSM-RN-All take much longer.


Therefore, we conclude that LSM-RN-Inc achieves a good trade-off between prediction accuracy and efficiency, which is applicable for real-time traffic prediction applications.

\vspace{-0mm}

\vspace{-0mm}

\vspace{-1mm}
\subsection{Varying parameters of our methods}
In the following, we test the performance of our methods by varying the parameters of our model. Since the effect of parameters does not have much correlation with the size of road network, we only show the experimental results on SMALL.

\vspace{-1mm}
\subsubsection{Effect of varying T}
\begin{figure}[!ht]
	\begin{small}\scalebox{0.99}{
			\begin{tabular}{cc}
				\vspace{-1mm} \\
				\hspace{-12mm}\includegraphics[width=50mm]{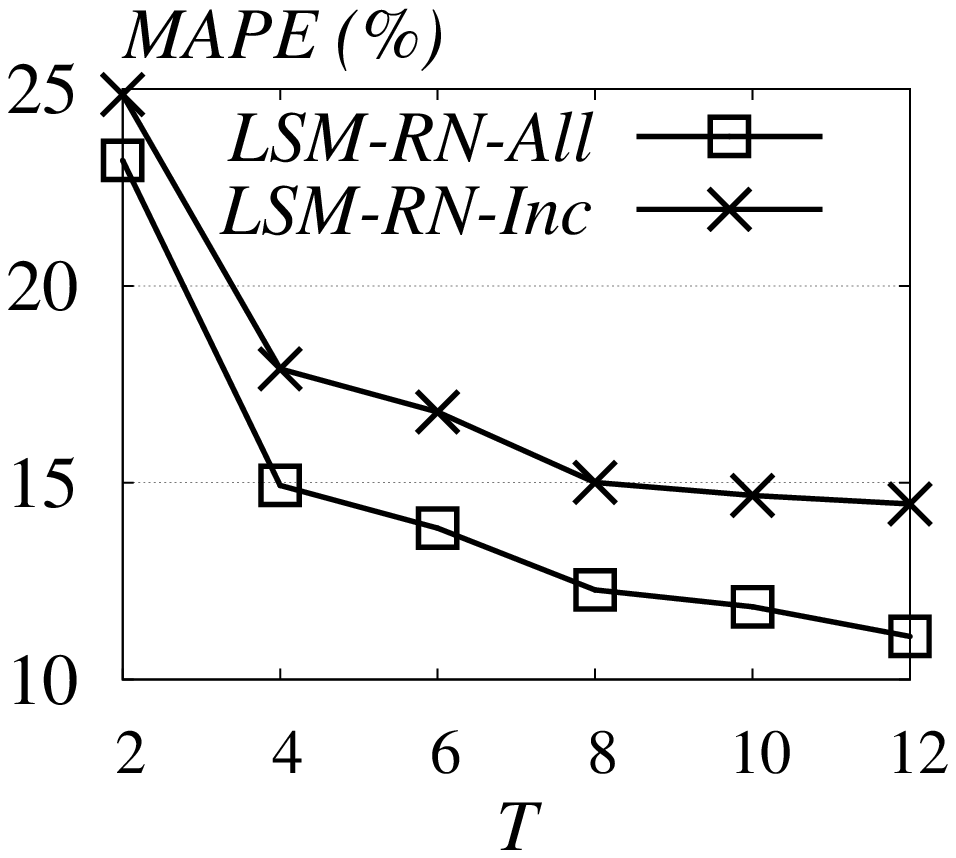} &
				\vspace{-2mm}\hspace{-12mm}\includegraphics[width=50mm]{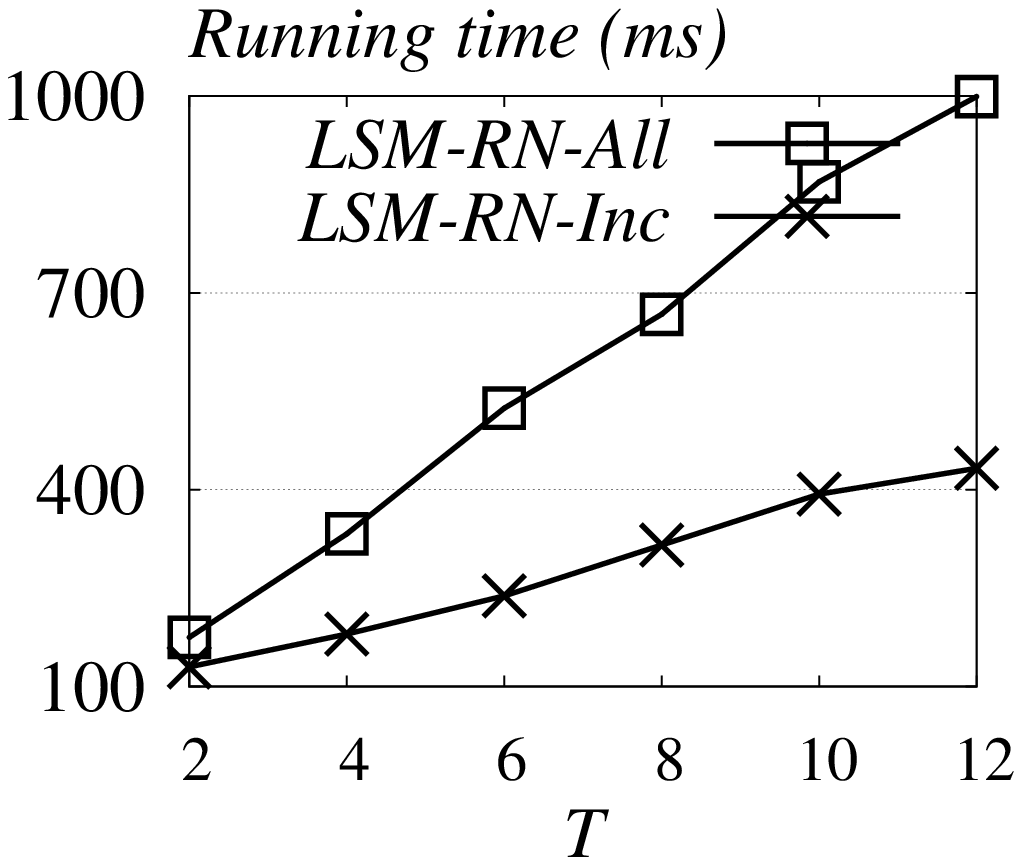}  \vspace{1mm}\\
				\hspace{-6mm}(a) Prediction error & \hspace{-5mm}(b) Running time
			\end{tabular}}
		\end{small}
		\vspace{-3mm}\caption{\small Effect of varying T} \vspace{-0mm}
		\label{fig::exp-varyT}
		\vspace{-2mm}
\end{figure}

Figure~\ref{fig::exp-varyT} (a) and Figure~\ref{fig::exp-varyT} (b) show the prediction performance and the running time of varying $T$, respectively.
We observe that with more number of snapshots, the prediction error decreases. In particular, when we increase $T$ from 2 to $6$, the results improve significantly. However, the performance tends to stay stable at $T\geq6$. This indicates that smaller number of snapshots (i.e., two or less) are not enough to capture the traffic patterns and the evolving changes. On the other hand, more snapshots (i.e., more historical data) do not necessarily yield much gain, considering the running time increases when we have more number of snapshots. Therefore, to achieve a good trade-off between running time and prediction accuracy, we suggest to use at least 6 snapshots, but no more than 12 snapshots.

\vspace{-0mm}
\begin{figure}[!ht]
	\begin{small}\scalebox{0.99}{
			\begin{tabular}{cc}
				\vspace{-1mm} \\
				\hspace{-12mm}\includegraphics[width=50mm]{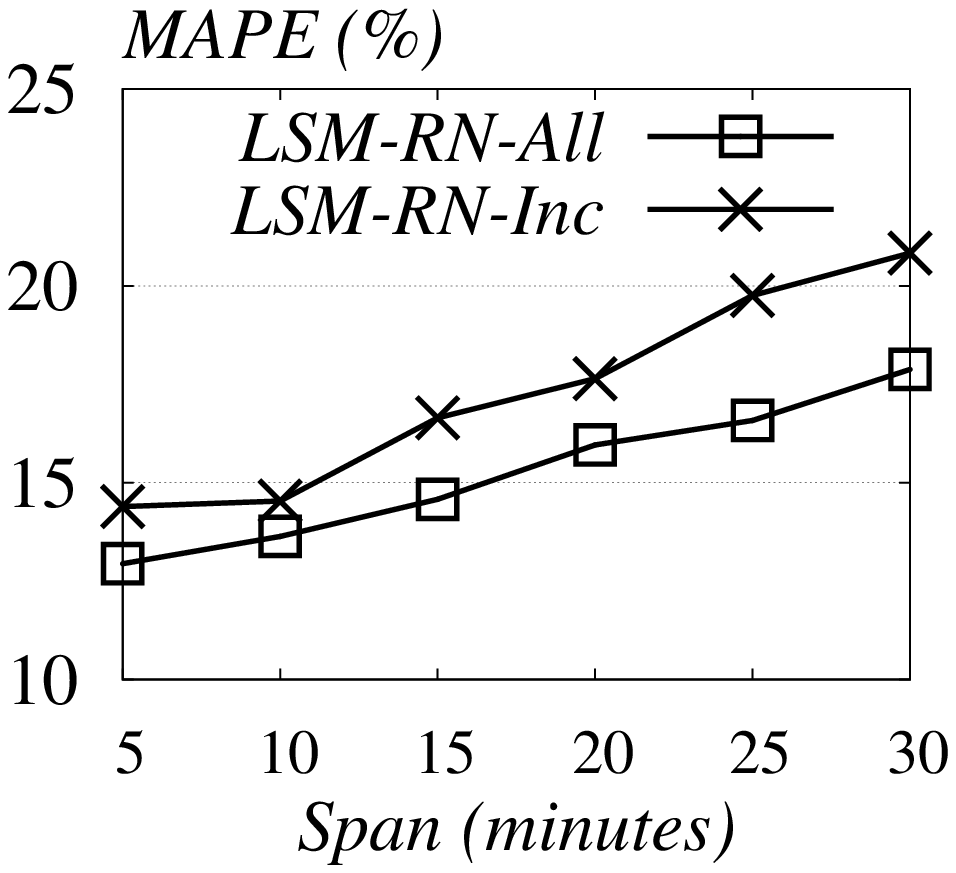} &
				\vspace{-2mm}\hspace{-12mm}\includegraphics[width=50mm]{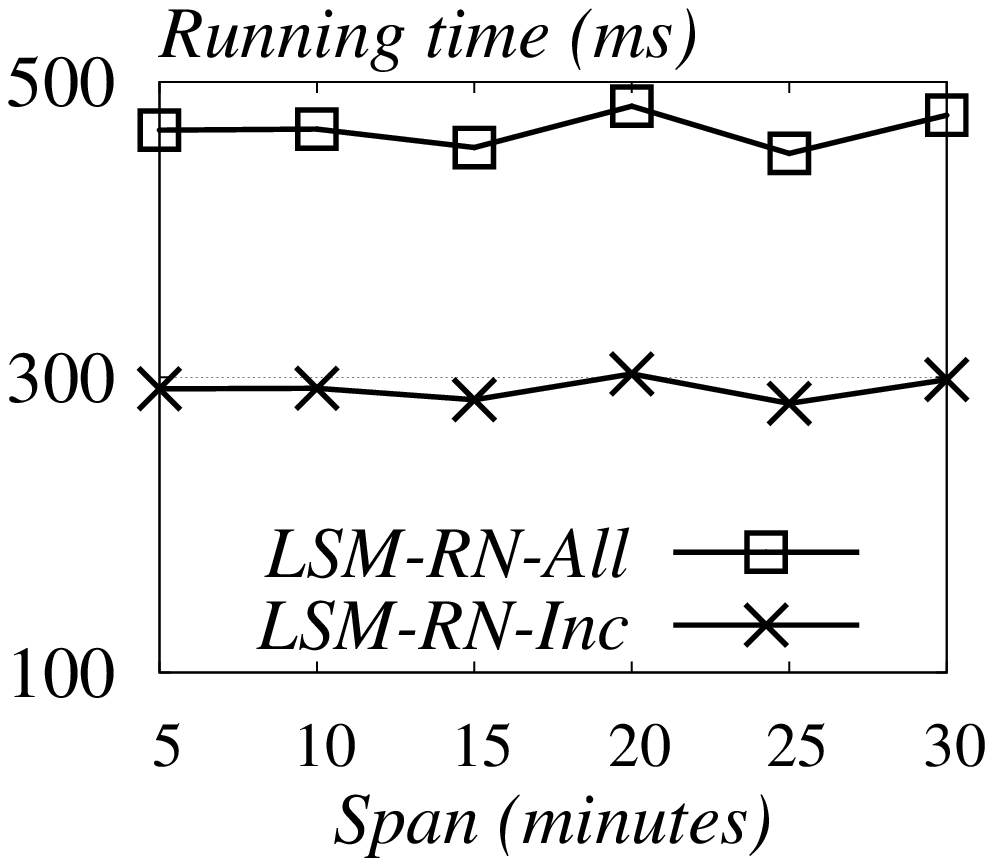}  \vspace{1mm}\\
				\hspace{-6mm}(a) Prediction error & \hspace{-5mm}(b) Running time
			\end{tabular}}
		\end{small}
		\vspace{-3mm}\caption{\small Effect of varying span} \vspace{-0mm}
		\label{fig::exp-varySpan}
		\vspace{-2mm}
	\end{figure}\vspace{-0mm}
	
	\begin{figure}[!ht]
		\begin{small}\scalebox{0.99}{
				\begin{tabular}{cc}
					\vspace{-1mm} \\
					\hspace{-12mm}\includegraphics[width=50mm]{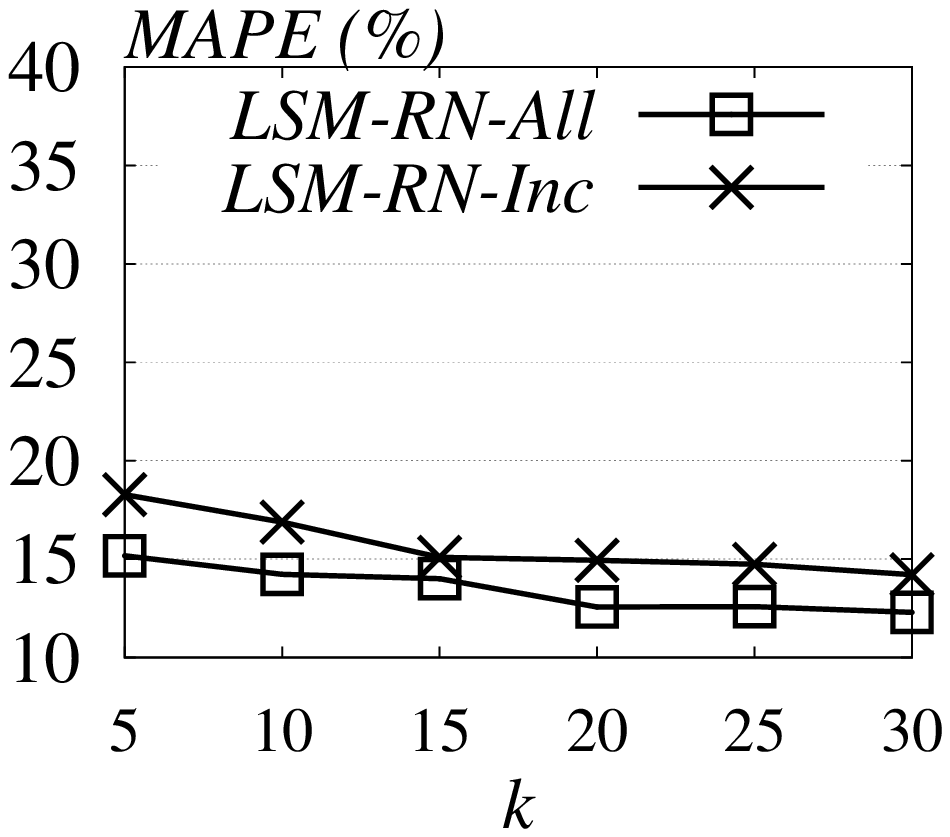} &
					\vspace{-2mm}\hspace{-12mm}\includegraphics[width=50mm]{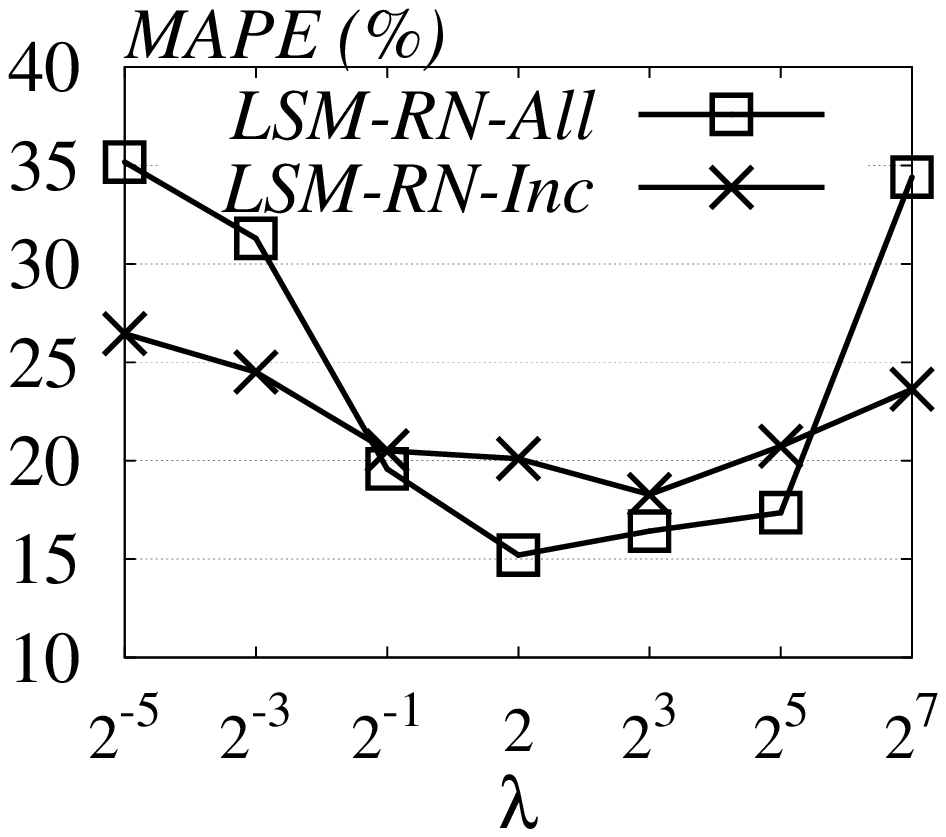}  \vspace{1mm}\\
					\hspace{-6mm}(a) Prediction MAPE with $k$ & \hspace{-5mm}(b) Prediction MAPE with $\lambda$
				\end{tabular}}
			\end{small}
			\vspace{-3mm}\caption{\small Effect of varying $k$ and $\lambda$ on prediction accuracy, where $k$ is number of latent attributes, and $\lambda$ is the graph regularization parameter. } \vspace{-0mm}
			\label{fig::exp-varyPara}
			\vspace{-2mm}
		\end{figure}\vspace{-0mm}

\subsubsection{Effect of varying span}
The results of varying $span$ are shown in Figure~\ref{fig::exp-varySpan}. It is clear that as the time gap between two snapshots increases, the performance declines. This is because when $span$ increases, the evolving process of underlying traffic may not stay smooth, the transition process learned in the previous snapshot are not applicable for the next prediction. Fortunately the sensor dataset usually have high-resolution, therefore it is always better to use smaller span to learn the latent attributes. In addition, span does not affect the running time of both algorithms.

\subsubsection{Effect of varying $k$ and $\lambda$}
Figure~\ref{fig::exp-varyPara} (a) shows the effect of varying $k$. We have two main observations from this experiment:  (1) we achieve better results with increasing number of latent attributes; (2) the performance stays stable when $k \geq 20$. This indicates a low-rank latent space representation can already capture the latent attributes of the traffic data. In addition, our results show that when the number of latent attributes is small (i.e., $k \leq 30$), the running time increased with $k$ but does not change much when we vary $k$ from 5 to 30. Therefore, setting $k$ as 20 achieves a good balance between computational cost and accuracy.


Figure~\ref{fig::exp-varyPara} (b) depicts the effect of varying $\lambda$, which is the regularization parameter for our graph Laplacian dynamics. We observe that the graph Laplacian has a larger impact on LSM-RN-All algorithm than that on LSM-RN-Inc. 
This is because $\lambda$ controls how the global structure similarity contributes to latent attributes and LSM-RN-All jointly learns those time-dependent latent attribute, thus $\lambda$ has larger effect on LSM-RN-ALL. 
In contrast, LSM-RN-Inc adaptively updates the latent positions of a small number of changed vertices in limited localized view, and thus is less sensitive to the global structure similarity than LSM-RN-ALL. In terms of parameters choices,  $\lambda=2$  and  $\lambda=8$ yields best results for LSM-RN-All and LSM-RN-Inc, respectively. 


\section{Conclusion}\label{sec-con}
In this paper, we studied the problem of real-time traffic prediction using real-world sensor data for road networks. We proposed LSM-RN, where each vertex is associated with a set of latent attributes that captures both topological and temporal properties of road networks.  We showed that the latent space modeling of road networks with time-dependent weights accurately estimates the traffic patterns and their evolution over time. To efficiently infer these time-dependent latent attributes, we developed an incremental online learning algorithm which enables real-time traffic prediction for large road networks. With extensive experiments we verified the effectiveness, flexibility and scalability of our model. For example, our incremental learning algorithm is orders of magnitude faster than the global learning algorithm, which takes less than 1 seconds to incorporate real-time feedback information for large road network.

For future work, we plan to embed the current framework into real applications such as ride-sharing or vehicle routing system, to enable better navigation using accurate time-dependent traffic patterns. Another interesting direction is to incorporate other data sources (e.g., GPS, incidents) for more accurate traffic prediction.




\bibliographystyle{abbrv}
\bibliography{traffic}

\section{Appendix}
\subsection{Derivatives of $L$ with respect to $U_t$ in equation~\ref{eqn::obj-L}.}\label{sec::app-1}
The objective of $L$ could be rewritten as follows:
\begin{equation}
\begin{aligned}
&L = J_1+J_2+J_3 + \sum_{t=1}^{T}Tr(\psi_tU_t)\\
&\mbox{where:}\\
&J_1=\sum_{t=1}^{T}Tr\Big(\big(Y_t\odot(G_t-U_tBU_t^T)\big)\big(Y_t\odot(G_t-U_tBU_t^T)\big)^T\Big) \\
&J_2=\sum_{t=1}^{T}\lambda Tr(U_tLU_t^T)\\
&J_3=\sum_{t=2}^{T}\gamma Tr\Big((U_t-U_{t-1}A)(U_t-U_{t-1}A)^T\Big)
\end{aligned}
\end{equation}

$J_1$ could also be rewritten as follows:
\begin{equation}
\begin{aligned} \label{eqn::j1}
J_1=&\sum_{t=1}^{T}Tr\Big(\big(Y_t\odot(G_t-U_tBU_t^T)\big)\big(Y_t\odot(G_t-U_tBU_t^T)\big)^T\Big)\\
=&\sum_{t=1}^{T}Tr\big( (Y_t\odot G_t)^T(Y_t\odot G_t)-2(Y_t^T\odot G_t^T)(Y_t\odot U_tBU_t^T)\\ & +(Y_t^T\odot U_tB^TU_t^T)(Y_t\odot U_tBU_t^T)\big)\\
=&\;const-2\sum_{t=1}^{T}Tr\big((Y_t^T\odot G_t^T)(Y_t\odot U_tBU_t^T)\big)\\&+\sum_{t=1}^T Tr\big((Y_t^T\odot U_tB^TU_t^T)(Y\odot U_tBU_t^T)\big)
\end{aligned}
\end{equation}

The second item of equation~\ref{eqn::j1} could be transformed by:
\begin{equation}
\begin{aligned} \label{eqn::rewritten}
O_1&=\sum_{t=1}^{T}Tr\big((Y_t^T\odot G_t^T)(Y_t\odot U_tBU_t^T)\big)\\
&=\sum_{t=1}^{T}\sum_{k=1}^{n}\big((Y_t^T\odot G_t^T)(Y_t\odot U_tBU_t^T)\big)_{kk}\\
&=\sum_{t=1}^{T}\sum_{k=1}^{n}\sum_{i=1}^{m}(Y_t^T\odot G_t^T)_{ki}(Y_t\odot U_tBU_t^T)_{ik}\\
&=\sum_{t=1}^{T}\sum_{i=1}^{m}\sum_{k=1}^{n}(Y_t\odot G_t \odot Y_t \odot U_tBU_t^T)_{ik}\\
&=\sum_{t=1}^{T}Tr\big((Y_t^T\odot G_t^T \odot Y_t^T) U_tBU_t^T\big)
\end{aligned}
\end{equation}

Now $J_1$ could be written as follows:
\begin{equation}
\begin{aligned} 
J_1=&\;const-2\sum_{t=1}^{T}Tr\big((Y_t^T\odot G_t^T \odot Y_t^T) U_tBU_t^T\big)\\&+\sum_{t=1}^T Tr\big((Y_t^T\odot U_tB^TU_t^T)(Y\odot U_tBU_t^T)\big)
\end{aligned}
\end{equation}

We now take the derivative of $L$ in respect of $U_t$:
\begin{small}
	\begin{equation}
	\begin{aligned}
	\frac{\partial L}{\partial U_t}=\frac{\partial J_1}{\partial U_t}+\frac{\partial J_2}{\partial U_t}+\frac{\partial J_3}{\partial U_t}+\frac{\partial \sum_{t=1}^{T}Tr(\psi_tU_t)}{\partial U_t}
	\end{aligned}
	\end{equation}
\end{small}

The derivative of  $\frac{\partial J_1}{\partial U_t}$ now could be calculated as follows:
\begin{small}
	\begin{equation}
	\begin{aligned}
	\frac{\partial J_1}{\partial U_t}= &-2(Y_t\odot G_t \odot Y_t)U_t B^T-2(Y_t^T\odot G_t^T \odot Y_t^T)U_t B\\&+\frac{\partial \sum_{t=1}^T Tr\big((Y_t^T\odot U_tB^TU_t^T)(Y\odot U_tBU_t^T)\big)}{\partial U_t}
	\end{aligned}
	\end{equation}
\end{small}

Suppose $O_2=\sum_{t=1}^T Tr\big((Y_t^T\odot U_tB^TU_t^T)(Y\odot U_tBU_t^T)\big)$, the derivative of $O_2$ could be written as:
\begin{equation}
\begin{aligned}
\frac{\partial O_2}{\partial U_t(pq)}&=\frac{\partial\sum_{k=1}^{n}\sum_{i=1}^{m}(Y_t^T\odot U_tB^TU_t^T)_{ki}(Y\odot U_tBU_t^T)_{ik}}{\partial U_t(pq)}\\
&=\frac{\partial\sum_{i=1}^{m}\sum_{k=1}^{n}(Y_t \odot Y_t \odot U_tBU_t^T\odot U_tBU_t^T)_{ik}}{\partial U_t(pq)}\\
&=\frac{\partial\sum_{i=1}^{m}\sum_{k=1}^{n}Y_t^2(ik)  (U_tBU_t^T)^2(ik)}{\partial U_t(pq)}
\end{aligned}
\end{equation}

Because only the $p_{th}$ row of $U_tBU_t^T$ is related with $\frac{\partial O_2}{\partial U_t(pq)}$, we have the following:
\begin{equation}
\begin{aligned}
\frac{\partial O_2}{\partial U_t(pq)}&=\frac{\partial\sum_{k=1}^{n}Y_t^2(pk)  (U_tBU_t^T)^2(pk)}{\partial U_t(pq)}\\
&=2\sum_{k=1}^{n}(Y_t^2)_{pk}(U_tBU_t^T)_{pk}\frac{\partial(U_tBU_t^T)_{pk}}{\partial (U_t)_{pq}}\\
&=2\sum_{k=1}^{n}(Y_t^2)_{pk}(U_tBU_t^T)_{pk}(U_tB^T+U_tB)_{kq}
\end{aligned}
\end{equation}

The matrices derivation is then expressed as:
\begin{equation}
\begin{aligned}
\frac{\partial O_2}{\partial U_t}&=2(Y_t\odot Y_t\odot U_tBU_t^T)(U_tB^T+U_tB)\\
&=2(Y_t\odot U_tBU_t^T)(U_tB^T+U_tB)
\end{aligned}
\end{equation}

Now the derivative of  $\frac{\partial J_1}{\partial U_t}$ is as follows:
\begin{small}
	\begin{equation}
	\begin{aligned}
	\frac{\partial J_1}{\partial U_t}= &-2(Y_t\odot G_t \odot Y_t)U_t B^T -2(Y_t^T\odot G_t^T \odot Y_t^T)U_t B\\&+2(Y_t\odot U_tBU_t^T\odot Y_t)(U_tB^T+U_tB)
	\end{aligned}
	\end{equation}
\end{small}

Similarly, we could calculate the derivatives of $\frac{\partial J_2}{\partial U_t}=2\lambda LU_t$,$\frac{\partial J_3}{\partial U_t}=2\gamma(U_t-U_{t-1}A) + 2\gamma(U_{t}AA^T-U_{t+1}A^T)$, and 
$\frac{\partial \sum_{t=1}^{T}Tr(\psi_tU_t^T)}{\partial U_t}=\psi_{t}$, we have the following:

\begin{equation}
\begin{aligned}
\hspace{-8mm}\frac{\partial L}{\partial U_t}= &-2(Y_t\odot G_t)U_t B^T-2(Y_t^T\odot G_t^T)U_t B+2(Y_t\odot U_tBU_t^T)(U_tB^T+U_tB)\\
&+2\lambda LU_{t}+ 2\gamma(U_t-U_{t-1}A) + 2\gamma(U_{t}AA^T-U_{t+1}A^T)+\psi_{t} \\
\end{aligned}
\end{equation}

\subsection{Update rule of $A$ and $B$}\label{sec::app-2}
Similar with the derivation of $U_t$, we add Lagrangian multiplier with $\phi \in R^{k\times k}$ and $\omega \in R^{k\times k}$, and calculate the derivatives of $L$ in respect of $A$ and $B$ :
\begin{equation}
\begin{aligned}
\frac{\partial L}{\partial B}&= -2\sum_{t=1}^{T}U_t^T(Y_t\odot G)U_t + 2\sum_{t=1}^{T}U_{t}^T(Y_t\odot U_{t}BU_{t}^T)U_{t}+\phi \\
\frac{\partial L}{\partial A}&=-2\sum_{t=2}^{T}U_{t-1}^TU_t+2\sum_{t=2}^{T}U_{t-1}^TU_{t-1}A+\omega \\
\end{aligned}
\end{equation}

Using the KKT conditions $\phi_{ij} B_{ij}=0$ and $\omega_{ij} A_{ij}=0$,
we get the following equations for $B_{kk}$, and $A_{kk}$:


\begin{equation}
\begin{aligned}
&-\big(\sum_{t=1}^{T}U_t^T(Y_t\odot G)U_t\big)_{ij}B_{ij} + \big(\sum_{t=1}^{T}U_{t}^T(Y_t\odot U_{t}BU_{t}^T)U_{t}\big)_{ij}B_{ij}=0
\end{aligned}
\end{equation}

\begin{equation}
\begin{aligned}
&-\big(\sum_{t=2}^{T}U_{t-1}^TU_t\big)_{ij}A_{ij}+\big(\sum_{t=2}^{T}U_{t-1}^TU_{t-1}A\big)_{ij}A_{ij}=0
\end{aligned}
\end{equation}
	
These lead us to the following update rules:
	

\begin{equation}
\begin{aligned}
&B_{ij}\leftarrow B_{ij} \Big(\frac{[\sum_{t=1}^{T}U_{t}^T(Y_t\odot G_{t}) U_{t}]_{ij}}{[\sum_{t=1}^{T}U_{t}^T(Y_t\odot (U_{t}BU_{t}^T))U_{t}]_{ij}}\Big)\\
\end{aligned}
\end{equation}

\begin{equation}
\begin{aligned}	
&A_{ij}\leftarrow A_{ij} \Big(\frac{[\sum_{t=1}^{T}U_{t-1}^TU_{t}]_{ij}}{[\sum_{t=1}^{T}U_{t-1}^TU_{t-1}A]_{ij}}\Big)\\
\end{aligned}
\end{equation}


%

\subsection{Extra experiment results based on RMSE}\label{sec::app-3}
In this set of experiments, we show the experiment results according to the measurement of RMSE, which indicates how closest predictions are to true observations. The definition of RMSE is as follows:

\begin{small}
\begin{equation*}
\begin{aligned}
RMSE = \sqrt{\frac{1}{N}\sum_{i=1}^{N}(y_i-\hat{y_i})^2}
\end{aligned}
\end{equation*}
\end{small}

\begin{figure}[!ht]
	\begin{small}\scalebox{0.99}{
			\begin{tabular}{cc}
				\multicolumn{2}{c}{\hspace{-5mm} \includegraphics[width = 82mm]{fig/Exp-Legends-comp}}
				\vspace{-1mm} \\
				\hspace{-12mm}\includegraphics[width=52mm]{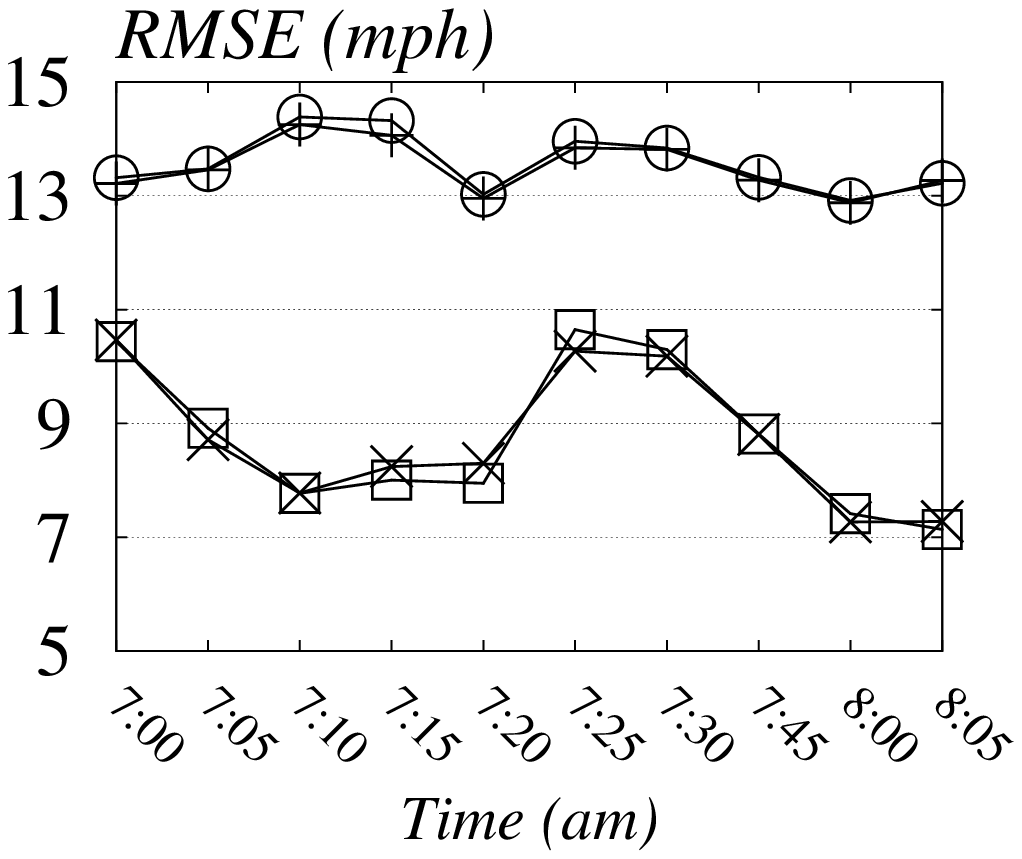} &
				\vspace{-2mm}\hspace{-12mm}\includegraphics[width=52mm]{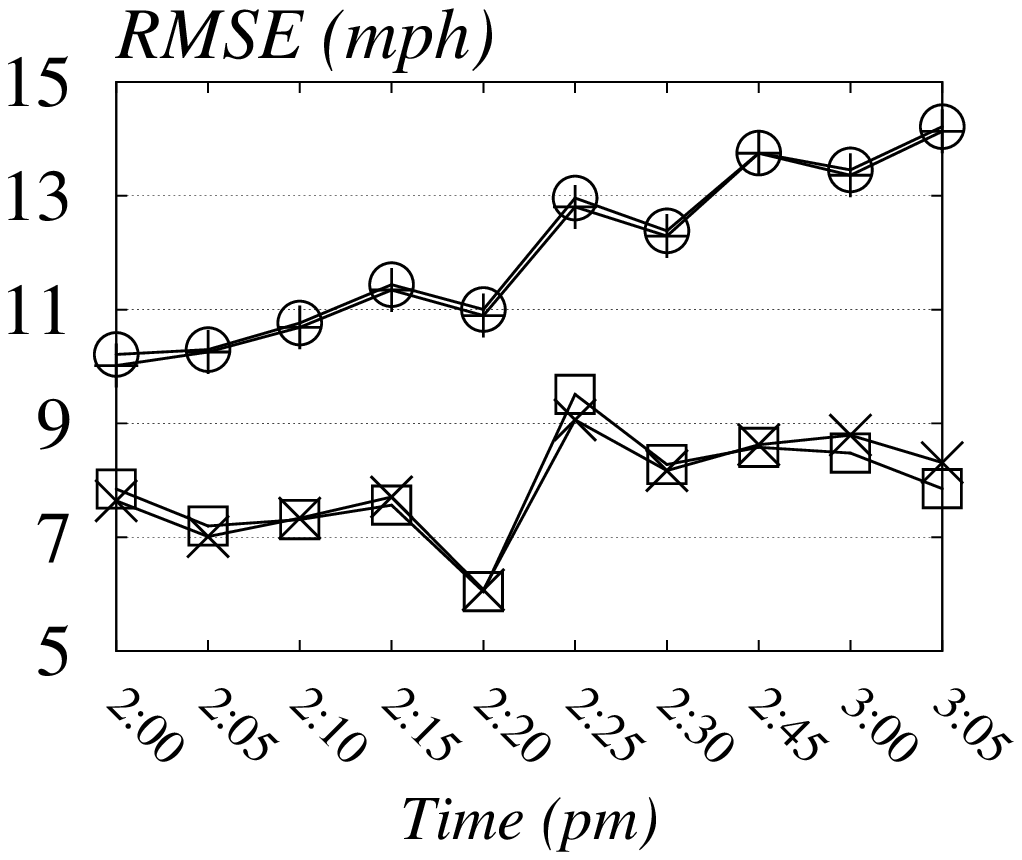}  \vspace{1mm}\\
				\hspace{-6mm}(a) Rush hour & \hspace{-5mm}(b) Non-Rush hour
			\end{tabular}}
		\end{small}
		\vspace{-3mm}\caption{\small Missing value completion rmse on SMALL} \vspace{-0mm}		
		\vspace{-0mm}\label{fig::exp-comp-rmse}
	\end{figure}

\begin{figure}[!ht]
		\begin{small}\scalebox{0.99}{
				\begin{tabular}{cc}
					\multicolumn{2}{c}{\hspace{-5mm} \includegraphics[width = 80mm]{fig/Exp-Legends-new}}
					\vspace{-1mm} \\
					\hspace{-12mm}\includegraphics[width=52mm]{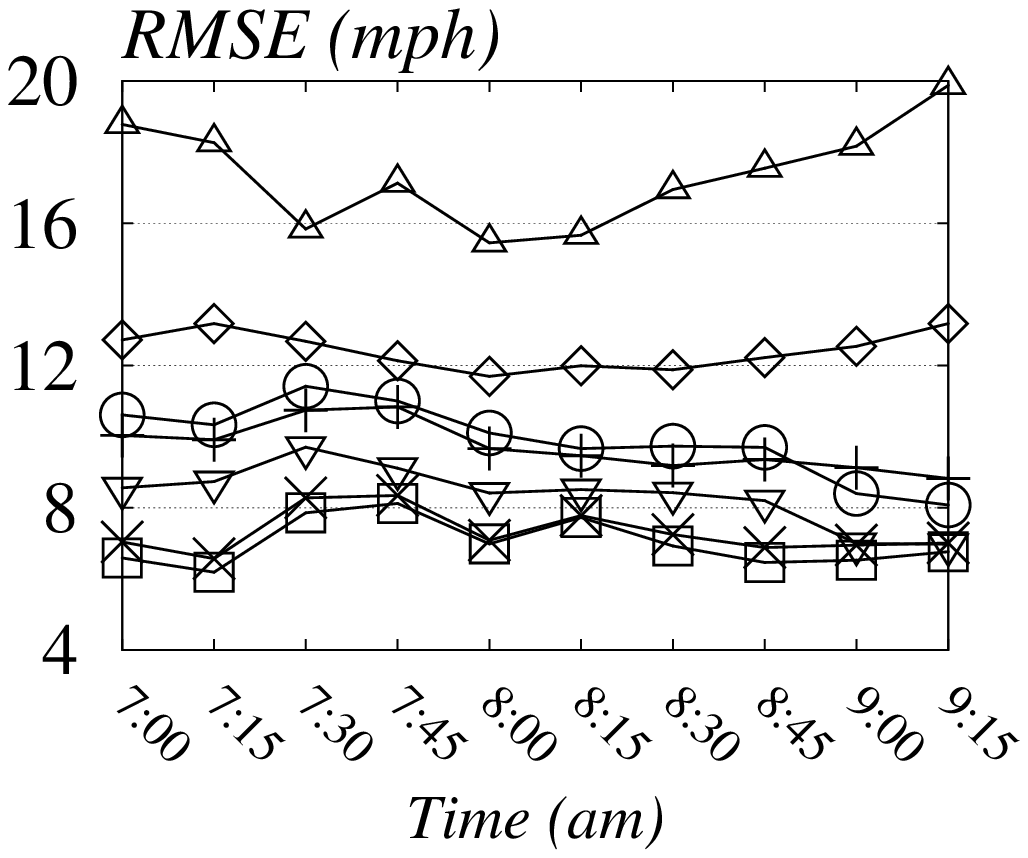} &
					\vspace{-2mm}\hspace{-12mm}\includegraphics[width=52mm]{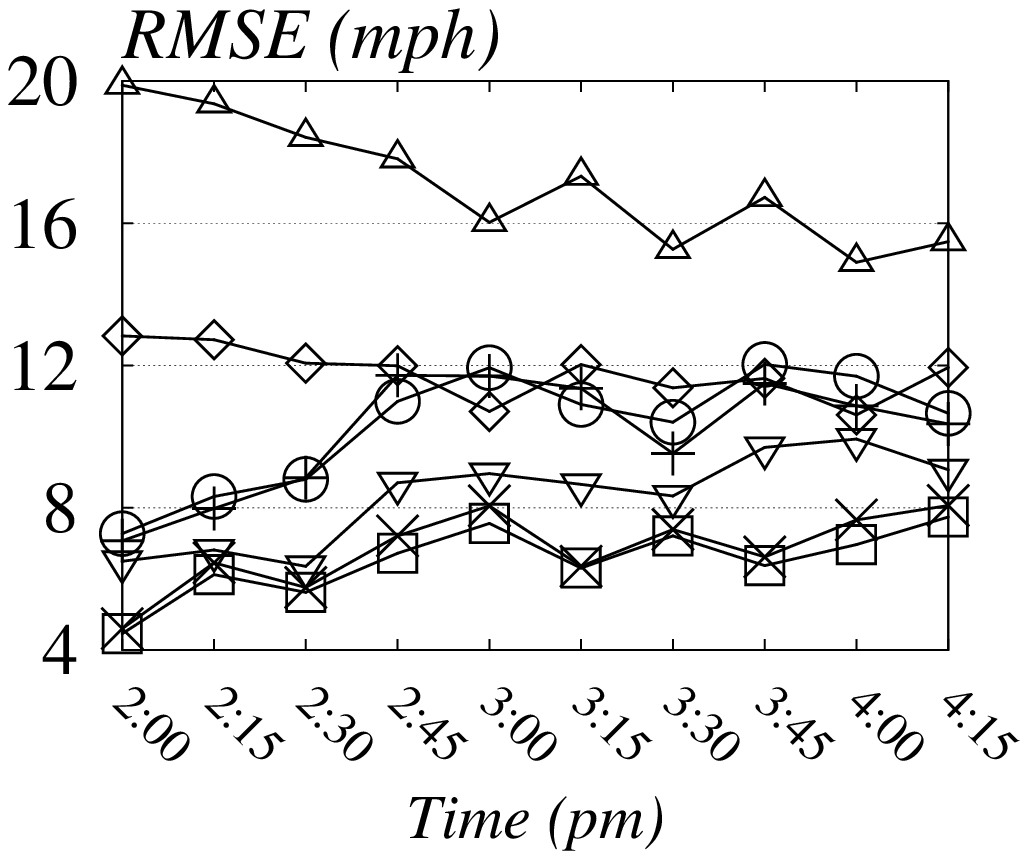}  \vspace{1mm}\\
					\hspace{-6mm}(a) Rush hour & \hspace{-5mm}(b) Non-Rush hour
				\end{tabular}}
			\end{small}
			\vspace{-3mm}\caption{\small Edge traffic prediction rmse one step on SMALL} \vspace{-0mm}
			\label{fig::exp-pred-rmse}
			\vspace{-0mm}
		\end{figure}

\begin{figure}[!ht]
	\begin{small}\scalebox{0.99}{
			\begin{tabular}{cc}
				\multicolumn{2}{c}{\hspace{-5mm} \includegraphics[width = 80mm]{fig/Exp-Legends-new}}
				\vspace{-1mm} \\
				\hspace{-12mm}\includegraphics[width=52mm]{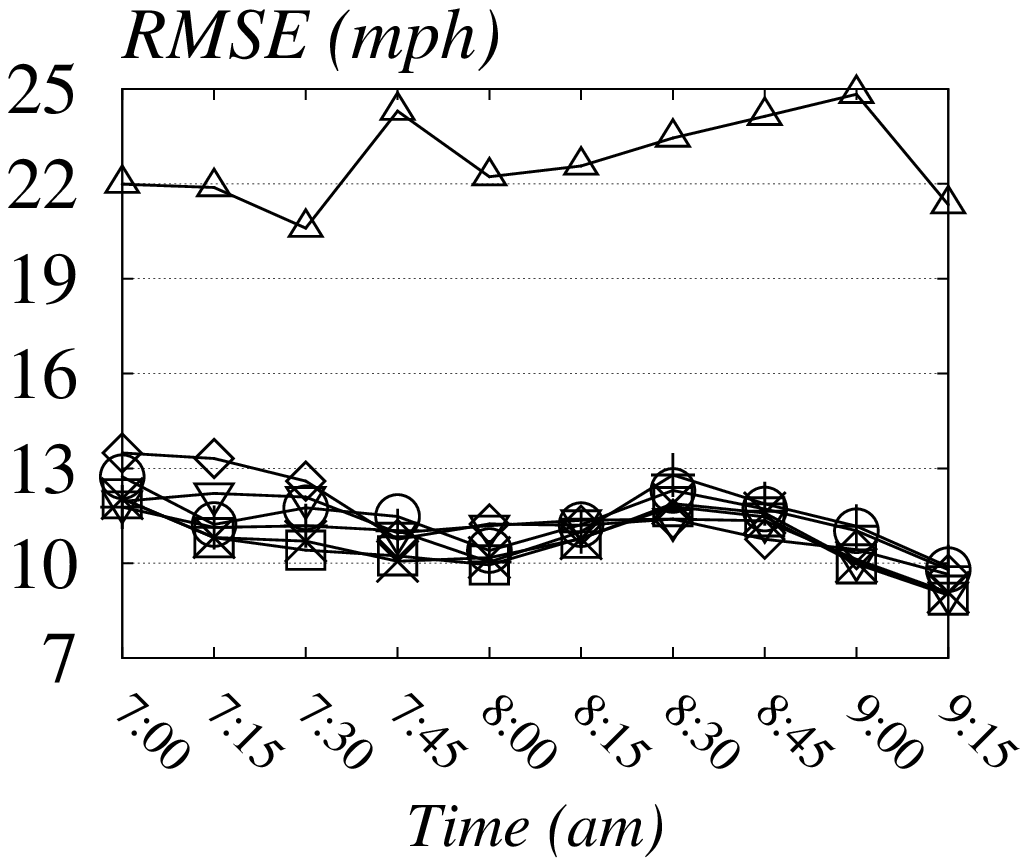} &
				\vspace{-2mm}\hspace{-12mm}\includegraphics[width=52mm]{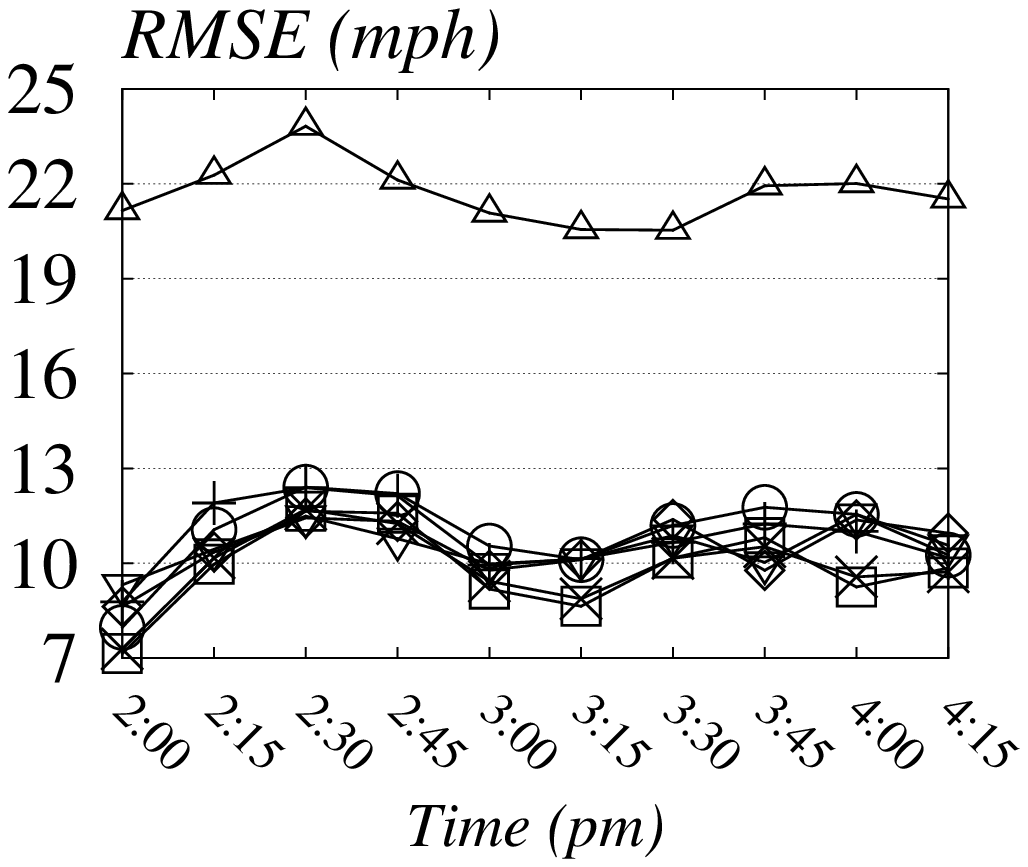}  \vspace{1mm}\\
				\hspace{-6mm}(a) Rush hour & \hspace{-5mm}(b) Non-Rush hour
			\end{tabular}}
		\end{small}
		\vspace{-3mm}\caption{\small Edge traffic prediction rmse 6 step ahead on SMALL} \vspace{-0mm}
		\label{fig::exp-pred-mult-rmse}
		\vspace{-0mm}
\end{figure}

\begin{figure}[!ht]
	\begin{small}\scalebox{0.99}{
			\begin{tabular}{cc}
				\multicolumn{2}{c}{\hspace{-5mm} \includegraphics[width = 82mm]{fig/Exp-Legends-ol-1}}
				\vspace{-1mm} \\
				\hspace{-12mm}\includegraphics[width=52mm]{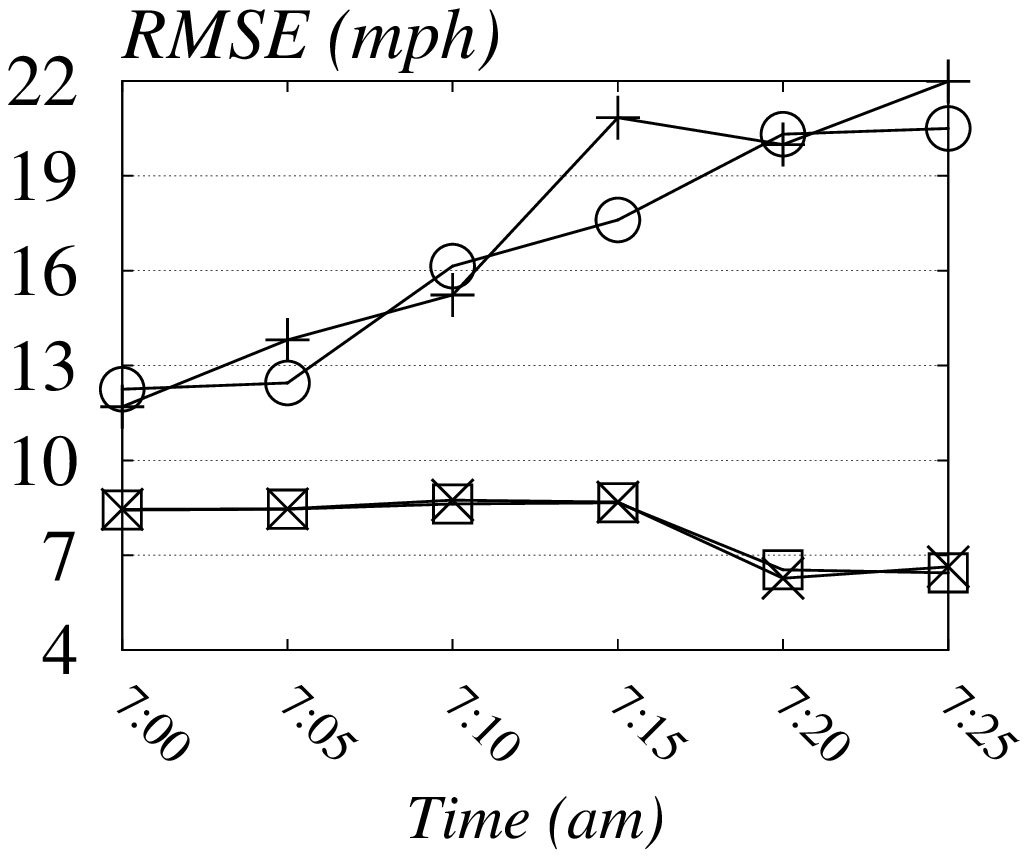} &
				\vspace{-2mm}\hspace{-12mm}\includegraphics[width=52mm]{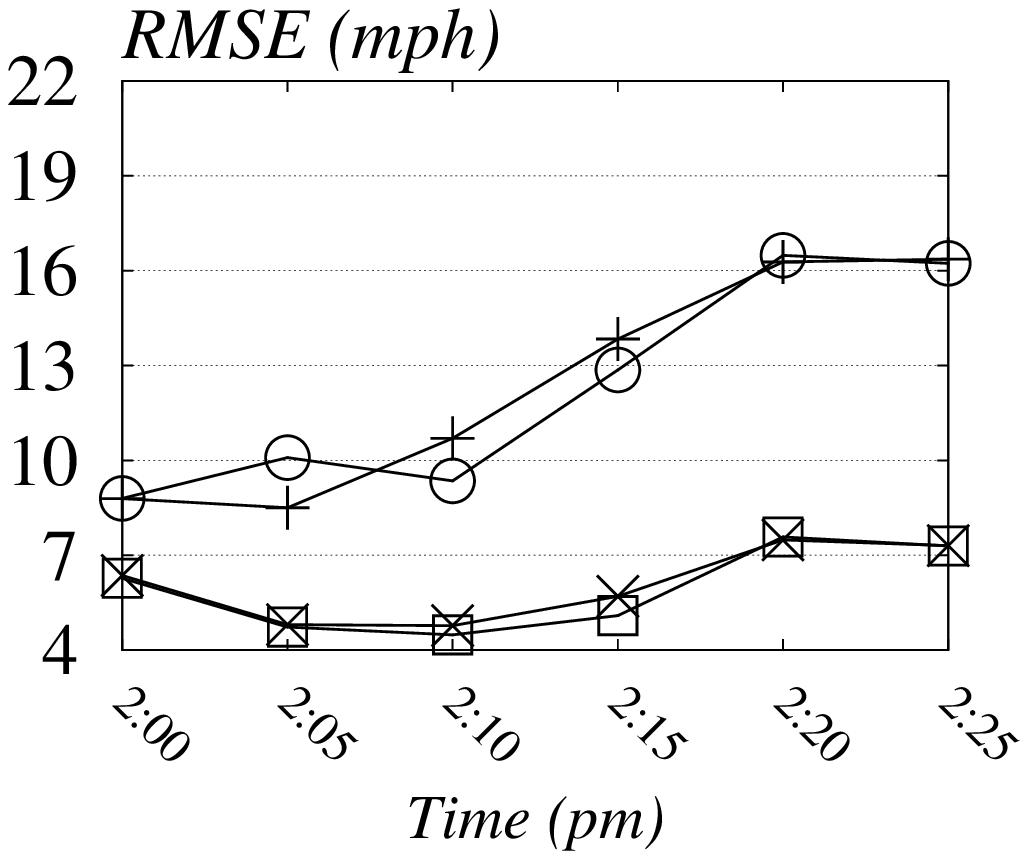}  \vspace{1mm}\\
				\hspace{-6mm}(a) Rush hour & \hspace{-5mm}(b) Non-Rush hour
			\end{tabular}}
		\end{small}
		\vspace{-3mm}\caption{\small Online Prediction rmse on SMALL} \vspace{-0mm}
		\label{fig::exp-online-rmse}
		\vspace{-0mm}
\end{figure}

Figure~\ref{fig::exp-comp-rmse} shows the experiment results for missing value completion. The one-step ahead prediction results are shown in Figure~\ref{fig::exp-pred-rmse}, six-step ahead prediction results are depicted in Figure~\ref{fig::exp-pred-mult-rmse}. Figure~\ref{fig::exp-online-rmse} shows the experiment results of online setting. The results based on RMSE are similar with those based on MAPE. 
We also observe that the predicated value by our methods deviate a small range (e.g., 4 to 11 mph) compared with the ground truth value.

\vspace{-0mm}
	
	\vspace{-0mm}
		
\vspace{-0mm}
	
\vspace{-0mm}

\vspace{-2mm}
\begin{figure}[!ht]
	\begin{small}\scalebox{0.99}{
			\begin{tabular}{cc}
				\vspace{-1mm} \\
				\hspace{-12mm}\includegraphics[width=52mm]{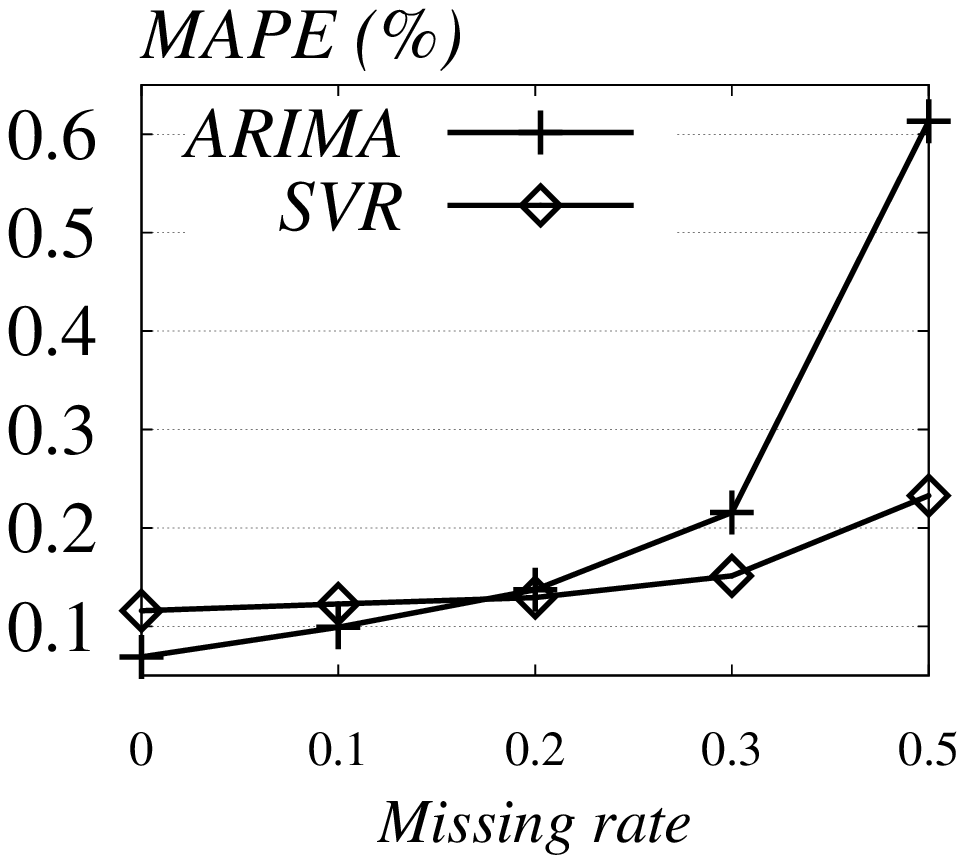} &
				\vspace{-2mm}\hspace{-12mm}\includegraphics[width=52mm]{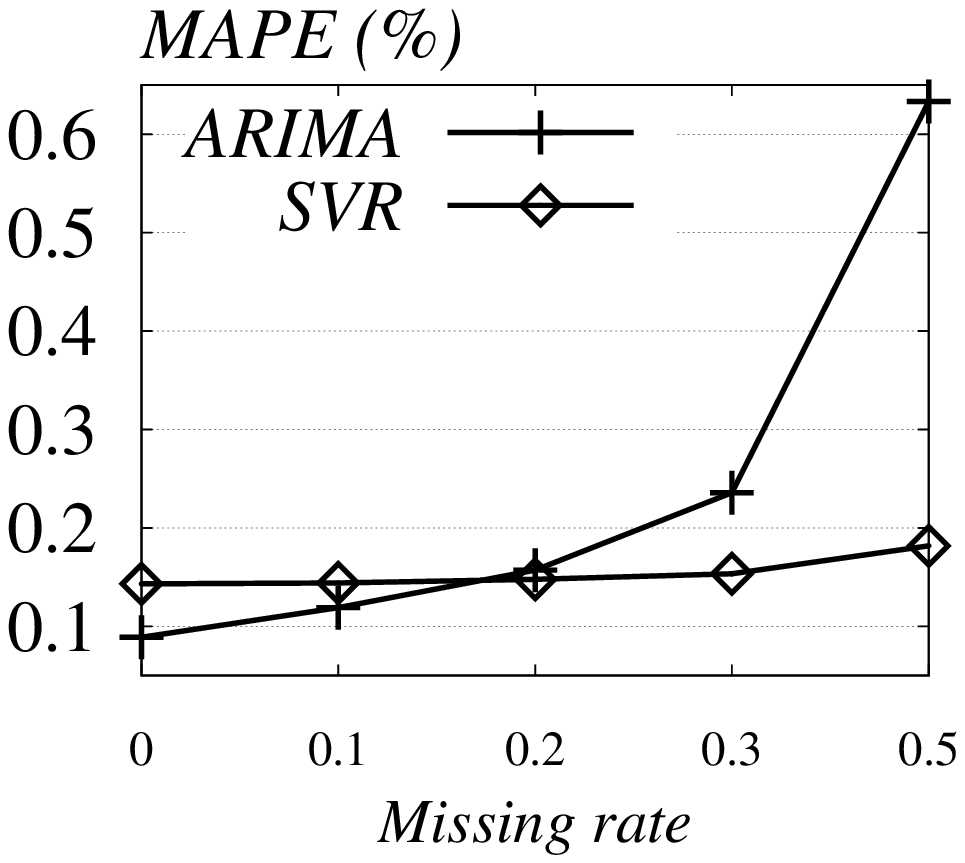}  \vspace{1mm}\\
				\hspace{-6mm}(a) SMALL & \hspace{-5mm}(b) LARGE
			\end{tabular}}
		\end{small}
		\vspace{-3mm}\caption{\small Missing rate during training stages for SVR and ARIMA} \vspace{-0mm}
		\label{fig::exp-miss-rate}
		\vspace{-0mm}
	\end{figure}\vspace{-0mm}

\vspace{-2mm}
\subsection{Effect of missing data}\label{sec::app-4}
In this set of experiment, we analyze the effect of missing data on the training dataset for the time series prediction techniques (i.e., ARIMA and SVR). The results are shown in Figure~\ref{fig::exp-miss-rate}. As shown in Figure~\ref{fig::exp-miss-rate} (a) and (b), the prediction error for both approaches increases with more number of noise. Similar to the effect of missing value on the prediction stages shown in Figure~\ref{fig::exp-pred-mape}, ARIMA is less robust than SVR because of its linear model. One interesting observation is that ARIMA performs better than SVR if the missing ratio is less than $10\%$, this indicates ARIMA is a good candidate for accurate traffic condition under the presence of complete data, this also conforms with the experiment results on~\cite{PanICDM12}. However, ARIMA is sensitive to the missing values during both training and prediction stages, which renders poor performance with incomplete dataset. 



\end{document}